\documentclass[12pt]{article}
 
\usepackage{setspace}
\usepackage[utf8]{inputenc}
\usepackage{graphicx, amsthm, amsbsy, amssymb, amsmath, natbib, bbm, color, fullpage, bm, booktabs}
\usepackage{caption}
\usepackage{subcaption}
\usepackage{rotating}
\usepackage[hidelinks]{hyperref}

\usepackage{xcolor}

\usepackage{xr}

\makeatletter
\newcommand*{\addFileDependency}[1]{% argument=file name and extension
\typeout{(#1)}
\@addtofilelist{#1}
\IfFileExists{#1}{}{\typeout{No file #1.}}
}
\makeatother

\newcommand*{\myexternaldocument}[1]{%
\externaldocument{#1}%
\addFileDependency{#1.tex}%
\addFileDependency{#1.aux}%
}

\myexternaldocument{supp_file}

\title{\vspace{-2cm} Nonparametric Bayesian Approach to Treatment Ranking in Network Meta-Analysis with Application to Comparisons of Antidepressants}
\author{Andr\'es F. Barrientos$^1$,  Garritt L. Page$^{2}$, and Lifeng Lin$^{1, 3}$}
\date{ 
\small
$^1$Department of Statistics, Florida State University\\%
$^2$Department of Statistics, Brigham Young University\\
$^3$Department of Epidemiology and Biostatistics, University of Arizona\\
\vspace{5mm}%
\large
\today
}
\begin{document}

\maketitle

\begin{abstract}
Network meta-analysis is a powerful tool to synthesize evidence from independent studies and compare multiple treatments simultaneously. A critical task of performing a network meta-analysis is to offer ranks of all available treatment options for a specific disease outcome. Frequently, the estimated treatment rankings are accompanied by a large amount of uncertainty, suffer from multiplicity issues, and rarely permit ties. These issues make interpreting rankings problematic as they are often treated as absolute metrics. To address these shortcomings, we formulate a ranking strategy that adapts to scenarios with high order uncertainty by producing more conservative results. This improves the interpretability while simultaneously accounting for multiple comparisons. To admit ties between treatment effects, we also develop a Bayesian Nonparametric approach for network meta-analysis. The approach capitalizes on the induced clustering mechanism of Bayesian Nonparametric methods producing a positive probability that two treatment effects are equal. We demonstrate the utility of the procedure through numerical experiments and a network meta-analysis designed to study antidepressant treatments.
\end{abstract}
{\bf Key words:} multiple comparisons, network meta-analysis, nonparametric Bayesian approach, stick-breaking process, spike and slab, treatment ranking. 

\doublespacing

\section{Introduction}
\label{sec:intro}

Network meta-analysis (NMA) is a powerful tool to synthesize evidence from independent studies and compare multiple treatments simultaneously \citep{lumley2002network, lu2004combination, salanti2012indirect}. It has been increasingly used in comparative effectiveness research and evidence-based medicine for making reliable medical decisions \citep{cipriani2013conceptual, mills2013demystifying, hutton2015prisma, riley2017multivariate}. Compared with conventional pairwise meta-analyses that rely on direct comparisons, an NMA synthesizes both direct and indirect evidence and thus typically produces more precise treatment effect estimates \citep{jackson2017borrowing, lin2019borrowing}.

An attractive feature of an NMA is that it can yield coherent rankings of all available treatments for a specific disease outcome \citep{higgins2015network}. Such rankings offer an intuitive way for clinicians to select optimal treatments. \cite{salanti2011graphical} proposed the surface under the cumulative ranking curve (SUCRA) to quantitatively summarize the performance of treatments in an NMA.\@ The SUCRA has been increasingly reported in recent NMAs for quantifying treatment rankings \citep{petropoulou2017bibliographic}, and it has been advocated in the reporting guideline of NMAs \citep{hutton2015prisma}. This measure ranges from 0 to 1; a higher value indicates a better treatment performance. Alternative ranking methods, such as the P-score, have been studied in the recent evidence synthesis literature \citep{rucker2015ranking, rucker2017resolve, mavridis2020extensions, rosenberger2021predictive}.

Despite their wide use in the current NMAs, the SUCRA and other existing treatment ranking measures have been criticized due to their potentially large uncertainties \citep{trinquart2016uncertainty}. For example, it is unclear whether the difference is substantial between a treatment with SUCRA=0.80 and another with SUCRA=0.75. Clinicians may misinterpret these treatment ranking measures as absolute metrics, and their uncertainties may be overlooked in clinical practice. Few studies have been devoted to account for uncertainties of treatment ranking measures \citep{veroniki2018providing, wu2021using}. 

Another critical problem of interpreting treatment rankings in an NMA is the multiplicity issue \citep{efthimiou2020dark}. This issue is due to the nature of NMAs: an NMA compares multiple treatments simultaneously, essentially leading to a problem of multiple hypothesis testings, while clinicians frequently interpret the effect estimate of each pair of treatment comparison separately, and thus the false positive rates could be seriously inflated. For example, consider a study with multiple treatments, labeled 1 to 10, that are compared to a placebo. Suppose that the effect is quantified by the mean difference, and all treatments actually have no effects compared with the placebo, so all mean differences have a true value of 0. We also assume that the mean differences independently follow standard normal distributions for convenience. In a conventional pairwise meta-analysis for treatment~1 vs.\ placebo, the probability of observing a mean difference of at least 1 is 15.9\%. If we consider the comparisons of treatment~2 vs.\ placebo and treatment~3 vs.\ placebo simultaneously, then the probability of observing the best treatment with a mean difference of at least 1 becomes 29.2\%. If we further consider all ten comparisons with placebo, then the probability of observing the best treatment with a mean difference of at least 1 becomes 82.2\%. As such, in an NMA with more treatment comparisons, we are more likely to observe exaggerated treatment effects. 

In addition, traditional Bayesian approaches for NMA do not provide tools that can be used to formally test equality between treatments. In an NMA, clinicians are first interested in determining that there is indeed a difference between treatments and, if so, then estimating the effect size. Traditional Bayesian approaches are not capable of testing for equality between treatments because the prior distributions employed do not put a positive probability on ties, that is, the event that two treatments are equal.

To overcome the foregoing limitations of existing NMA methods, we develop an ordering method based on ideas that are similar in spirit to that by \cite{barrientos2021} that accounts for multiplicity and develops a Bayesian nonparametric (BNP) approach that allows ties with positive probability. Our approach orders treatment effects so that order uncertainty and multiple comparisons are deliberately considered. This is done by sequentially identifying ordering sets that have high posterior probabilities. As a result, order uncertainty is rationally included by remaining silent about the order of specific units when the uncertainty associated with their treatment estimates is high. When comparing treatment effects, we not only focus on whether a treatment effect is greater or less than another one but also on whether they are equal (i.e., ties are permitted). This requires a prior that places a positive probability on the event that two treatments have equal effects. For this reason, we develop a novel BNP approach to modeling treatment effects in an NMA\@. Our BNP model employs a spike and slab base measure \citep{kim&dahl&vannucci:2009, canale:2017}, which permits us to easily incorporate ties among treatments in our rankings. 

This article is organized as follows. Section~\ref{sec:data} describes an NMA dataset of antidepressant comparisons originally reported by \cite{cipriani2009comparative}, which will be used to illustrate the structure of NMA and serve as a real-world data analysis. Section~\ref{sec.models} details a commonly-used model for performing an NMA. This article will consider the contrast-based framework that focuses on estimating treatment contrasts, while the literature also has extensive discussions on the use of arm-based NMA models \citep{dias2016absolute, hong2016rejoinder, lin2017performing, white_etal:2019}. Section~\ref{sec.ordering.statements} describes our approach to ordering treatment statements in a traditional NMA setting. This motivates proposing alternative models, which are described in Section~\ref{sec:newmodels}. We then compare the performance from our new method to that of existing methods using simulation studies in Section~\ref{sec:sim} and, in Section~\ref{sec:realdata}, we compare methods using the antidepressants dataset. Finally, this article concludes with a brief discussion in Section~\ref{sec:disc}.

\section{Preliminaries}
\subsection{Description of Antidepressants Dataset}
\label{sec:data}

We use a dataset of antidepressants, originally reported by \cite{cipriani2009comparative}, to illustrate the conventional NMA models and our proposed models. The dataset is comprised of 111 randomized controlled trials on major depression with 12 unique drug treatments. The original systematic review considered both the efficacy (reflected by response rate) and acceptability (reflected by dropout rate) outcomes. This article is restricted to NMA methods for a single outcome, instead of jointly modeling multiple outcomes; we focused on the efficacy outcome, which is binary. In our analysis, we label the 12 treatments in alphabetical order: 1) bupropion; 2) citalopram; 3) duloxetine; 4) escitalopram; 5) fluoxetine; 6) fluvoxamine; 7) milnacipran; 8) mirtazapine; 9) paroxetine; 10) reboxetine; 11) sertraline; and 12) venlafaxine. The odds ratio is used as the effect measure. For the efficacy outcome, the odds ratio of A vs.\ B higher than 1 indicates that treatment~A performs better than treatment~B\@.

\begin{figure}
\centering
\includegraphics[width = .5\textwidth]{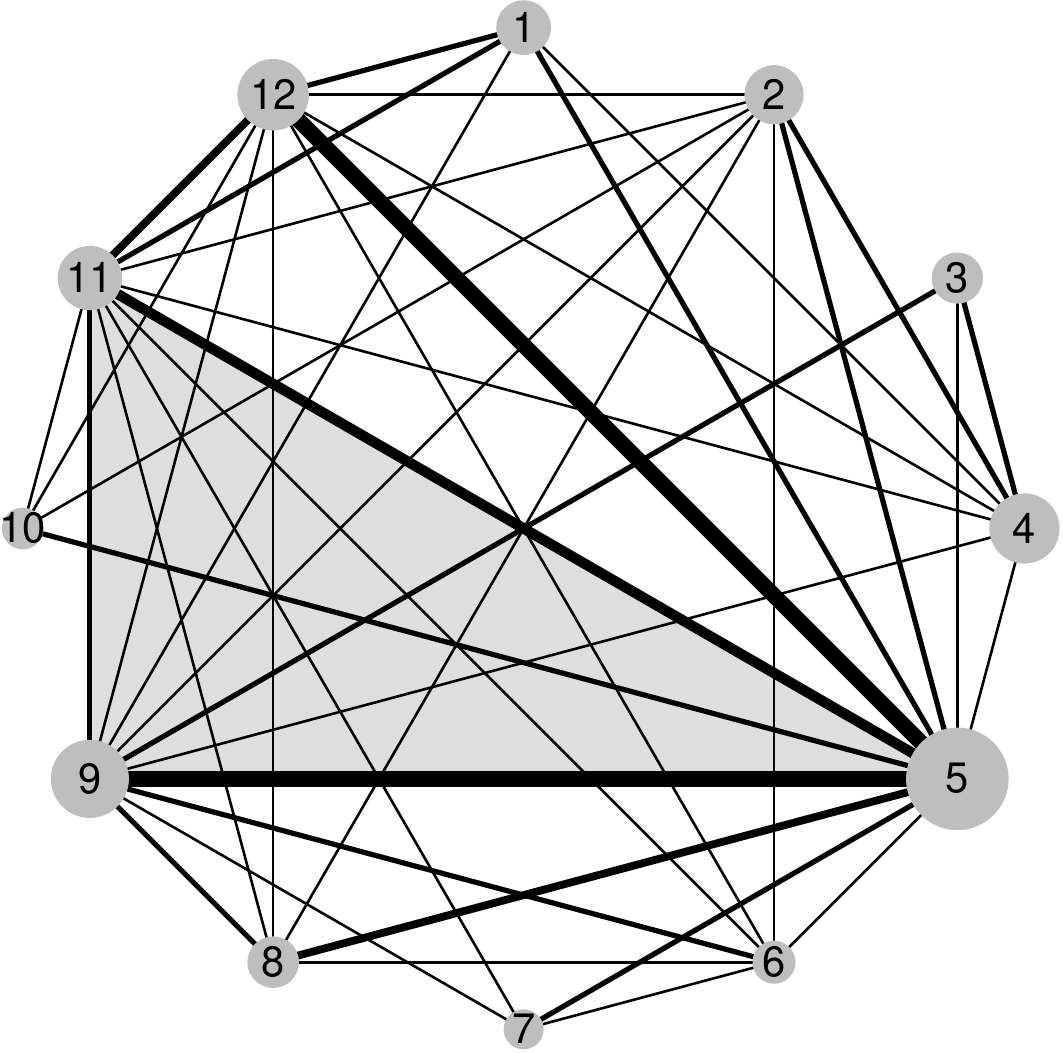}
\caption{Network plot of the dataset of antidepressants. Each node represents an antidepressant, and each edge represents a direct comparison between the corresponding two antidepressants. The node size is proportional to the total sample size in the corresponding groups of antidepressants across studies, and the edge width is proportional to the number of studies that directly compare the corresponding antidepressants. The shaded loop represents the three-arm studies.}
\label{fig:adnetwork}
\end{figure}

Figure~\ref{fig:adnetwork} presents the network plot of this dataset. All studies but two were two-arm trials, while the other two studies were three-arm trials, comparing treatments~5, 9, and 11. Fluoxetine (treatment~5) had been conventionally used as a reference antidepressant \citep{magni2013fluoxetine}; as such, it involves the largest sample size and the largest number of trials, as indicated in the network plot. This network is generally well connected, with many direct comparisons for treatment pairs. Nevertheless, most pairs of treatments are directly informed by a limited number of trials, and some comparisons (e.g., treatments~1 vs.\ 2 and 2 vs.\ 3) lack direct evidence. The NMA approach provides a valuable way for synthesizing direct and indirect comparisons in the same analysis and offering a coherence ranking of all available antidepressants.

\subsection{Contrast-Based Model for Network Meta-Analysis}
\label{sec.models}

An NMA is commonly carried out using either a contrast-based or arm-based approach \citep{zhang2014network, efthimiou2016getreal}. The key difference between these two approaches is that the contrast-based approach assumes that relative treatment effects are exchangeable across studies, while the arm-based approach assumes exchangeability across studies for absolute treatment effects. \cite{white_etal:2019} offer a detailed comparison between the two approaches. Our focus is the contrast-based approach because assuming exchangeability across studies for the relative treatment effects is more widely used in current evidence synthesis \citep{dias2016absolute}.

For context, we detail the contrast-based NMA model used by \cite{lu2004combination, lu&ades:2006} for binary outcomes, as in the antidepressant dataset in Section~\ref{sec:data}. Let $N$ be the total number of studies and $K$ be the total number of unique treatments across all studies. We use $\mathcal{T}_i$ to denote the collection of treatments that are included in study~$i$ ($i = 1, \ldots, N$). Further, $y_{ik}$ denotes the number of event outcomes (``successes'') associated with treatment $k \in \mathcal{T}_i$ and $n_{ik}$ the number of trials. Next, for study~$i$, we use $b_i \in \mathcal{T}_i$ to denote the baseline treatment while $\mathcal{T}^{-b_i}_i = \mathcal{T}_i \setminus \{b_i\}$ denotes the collection of treatments excluding the baseline treatment. With this notation, consider the following random-effects model:
\begin{align} 
    y_{ik}|p_{ik} & \overset{\text{ind}}{\sim} \mbox{Binomial}(n_{ik},p_{ik}) \ \mbox{for $i = 1, \ldots, N$ and $k \in \mathcal{T}_i$}; \label{likelihood} \\ 
    {\rm logit}(p_{ik}) & = \mu_{i,b_i} + \delta_{i, b_ik}\mathbb{I}(b_i \neq k) \ k \in \mathcal{T}_i; \\
    \mu_{i, b_i} & \overset{\text{iid}} \sim \mbox{N}(m_b,s_b);\\
    \delta_{i, b_ik} | d_{1b_i}, d_{1k} & \overset{\text{ind}}{\sim} \mbox{N}( d_{1k} - d_{1b_i}, \tau), \ k \in \mathcal{T}^{-b_i}_i; \label{delta_model}\\
    {\rm Corr}(\delta_{i, b_ik}, \delta_{i, b_ih}) & = \gamma=0.5, \ k, h \in \mathcal{T}^{-b_i}_i, k \ne h; \label{delta_correlation}\\
     \tau^2 & \sim \mbox{LogN}(m_{\ell}, s_{\ell}); \label{tau_prior}\\
%    \tau & \sim \mbox{uniform}(0,m_{\tau}) \\
    d_{1k}  & \stackrel{\text{iid}}{\sim} \mbox{N}(m_d, s_d), \ k = 2, \ldots, K \ \mbox{with $d_{11} = 0$}.\label{lastprior}
%    d_{1k} & \sim N(0, 100), \ k = 2, \ldots, K \ \mbox{with $d_{11} = 0$}.\\
\end{align}
In this model, $p_{ik}$ represents the probability of success for treatment~$k$ in study~$i$, $\mu_{i, b_i}$ represents study~$i$'s baseline treatment effect, $\delta_{i, b_ik}$ represents the effect of treatment~$k$ to treatment $b_i$, $d_{1k}$ represents the mean effect of treatment~$k$ vs.\ treatment 1, the reference treatment. Of note, the baseline treatments and the reference treatment are not necessarily the same. If there are more than two treatments in a particular study (i.e., a multi-arm study), then the treatment effects are correlated. The correlation coefficient is typically assumed to be 0.5 \citep{higgins1996borrowing, lu2004combination}. All roman letters correspond to prior parameters that are user-supplied. The model in Equations~\eqref{likelihood}--\eqref{lastprior} is adopted from \cite{lu&ades:2006}, except they employ a uniform prior distribution for $\tau$. The log-Gaussian prior for $\tau^2$ in Equation~\eqref{tau_prior} is an alternative informative prior that is studied in \cite{turner_etal:2012}. In what follows, we will refer to this model as the ``Gaussian effects'' model because the $d_{1k}$ parameters are modeled using a Gaussian distribution. 

The Gaussian effects model assumes that direct evidence is consistent with indirect evidence. Specifically, the NMA synthesizes both direct evidence (comparisons of treatments that are within a study) and indirect evidence (comparisons of treatments that do not appear together in a study but have a common comparator). Comparisons between treatments and the reference (i.e., $d_{1k}$) are called ``basic parameters,'' while those that are linear combinations of the basic parameters (e.g., $d_{kk'} = d_{1k'} - d_{1k}$) are called ``functional parameters''. Functional parameters must be able to be written in terms of basic parameters. Under the assumption of evidence consistency, both direct and indirect comparisons estimate the same parameter.

%The model described in \eqref{likelihood} - \eqref{lastprior} has, in some ways, been extensively used when carrying out Bayesian network analysis from an effects perspective (\textcolor{red}{puts some cites}). The model employed for $d_{1k}$ \eqref{lastprior} can be limiting in the types of comparisons between treatments that are available. For example, it is not possible to infer if one treatment is exactly equal to another or if a treatment effect is exactly equal to 0. Both of of these types of inferences are available in our approach. We describe the extensions we make to \cite{lu&ades:2006}'s model and highlight how they enable making probabilistic statements with regards to two treatment effects being equal and/or a treatment effect being equal to 0. But first we provide some background.

\section{Treatment Effect Comparisons Using Directed Graphs}
\label{sec.ordering.statements}

Once the Gaussian effects model is fit, inference regarding treatment effect order is often desired. The pairwise comparisons available from the Gaussian effects model facilitate ordering the treatment effects. However, the ordering inherits the potentially high uncertainty that may exist among a subset of pairwise comparisons. This problem is not explicitly addressed by traditional methods that summarize treatment comparisons in an NMA \citep{trinquart2016uncertainty}. In particular, the issues that traditional methods do not simultaneously address are the following: i) multiplicity (i.e., complications that accompany inference in the case that all comparisons co-occur simultaneously), ii) identifying comparisons that have low uncertainty, and iii) representing all comparisons simultaneously in a concise and interpretable way. We develop a novel approach that addresses all three issues. Our approach seeks to identify treatments or groups of treatments whose comparisons are accompanied by low uncertainty (or high posterior probability). We account for multiple treatment comparisons by attaching a single posterior probability that comparisons occur simultaneously. Then, we display the simultaneous comparisons using an easy-to-interpret network graph from which treatment rankings can be extracted.

% What would mallow's model do?

% The goal is to provide a methodological approach that allows comparing treatment effects.

\begin{figure}[t]
    \centering
    \includegraphics[width =.45\textwidth, page=2]{Illustration_Network_Treatment.pdf}
    \includegraphics[width =.45\textwidth]{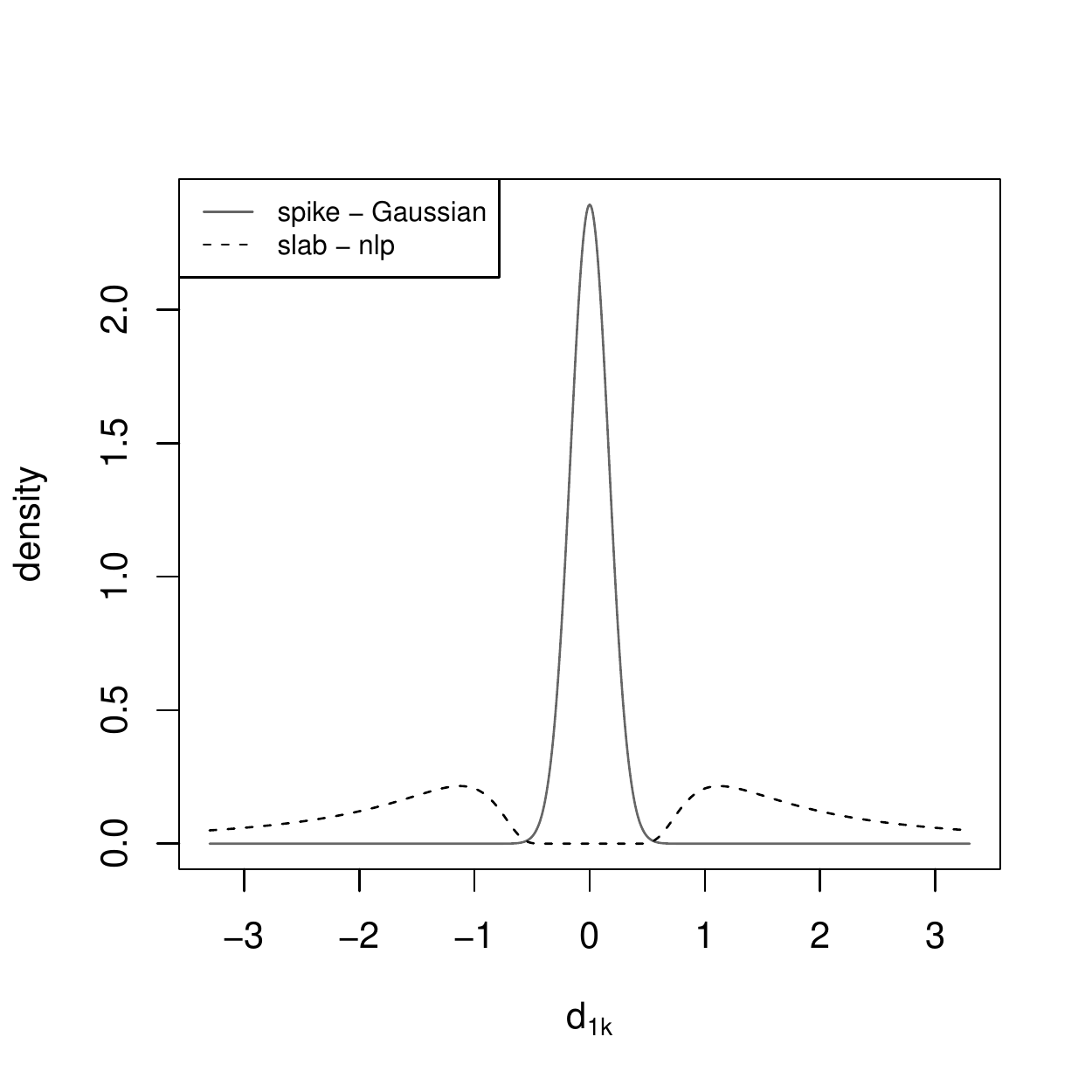}
    \caption{Illustrative plots. The left plot illustrates a network treatment effect when $K=7$ and $E = \{(1,2), (2,4), (4,2), (1,4), (1,6), (2,6), (3,7), (7,3)\}$. A dashed one-directional arrow from A to B indicates that treatment~B performs better treatment~A\@. A solid bi-directional arrow between A and B indicates that treatments~A and B have the same effect.  The right plot illustrates the spike and slab density where the spike is $N(0, v_0/3)$ and the slab is $\mbox{NLP}(p, 1, 1)$. The $p$ is selected so that $\mbox{NLP}(x_0; p, 1, 1) = \mbox{N}(x_0 | 0, v_0/3)$, where $x_0$ satisfies $\int_{-x_0}^{x_0}\mbox{N}(x | 0, v_0/3)dx \approx 0.999$. }
    \label{fig:Illust_Network_Treatment}
\end{figure}

Our approach summarizes the treatment effects in an NMA using a graphical structure. The nodes of a graph represent the $K$ treatments which are denoted by $ \mathcal{T} = \{1,\ldots,K\}$. The graphs considered are directed graphs with random edges. We use $\mathcal{E}$ to denote the random graph. The domain of $\mathcal{E}$ is the collection of all graphs 
% that are acyclic 
with edges representing a set of ordered pairs of nodes which in turn represent comparisons among the treatments. In other words,
\begin{itemize}
    \item[-] if $(j,k), (k,j) \in \mathcal{E}$, then  $d_{1j} = d_{1k}$;
    \item[-] if $(j,k)\in \mathcal{E}$ but $(k,j) \not\in \mathcal{E}$, then $d_{1j} < d_{1k}$.
\end{itemize}
As an example, let $K=7$ with $E = \{(1,2), (2,4), (4,2), (1,4), (1,6), (2,6), (3,7), (7,3)\}$ which results in
$$
\Pr(\mathcal{E} = E |  {\rm Data})
= 
\Pr(d_{11} < d_{12} < d_{16}, d_{11} < d_{14}, d_{12} = d_{14},   d_{13} = d_{17}|  {\rm Data}).
$$
Figure~\ref{fig:Illust_Network_Treatment} displays the graph $(\mathcal{T},E)$ described above, where dashed one-directional arrows point toward the treatment with larger effect. For example, treatment~6 has a larger effect compared to treatment~2. Solid bi-directional arrows represent the case where two treatments have the same effect. For example, treatments~2 and 4 have the same effect as well as treatments~3 and 7. A pair of nodes without an arrow connecting them indicates that no statement is made between them; in our strategy, no arrow between two nodes implies there is considerable uncertainty when comparing the two treatments. The kind of treatment rankings that can be extracted from the directed graph in Figure~\ref{fig:Illust_Network_Treatment} are that treatment 6 performs better than treatments 1 and 2, treatment 2 performs better than 1, treatments 2 and 4 have similar effects, and comparing treatment 6 to treatment 4 is inconclusive.

Our goal is to find graphs like those displayed in Figure~\ref{fig:Illust_Network_Treatment} that accumulate high posterior probability. This is accomplished by first creating a graph in which all nodes are connected through the relationship (e.g., $d_{1j} = d_{1k}$) that has the highest posterior probability. Jointly, the pairwise comparisons represented by this graph will typically have a low posterior probability. We then begin trimming the graph by removing arrows associated with relations that have the lowest posterior probability. This produces a sequence of graphs with varying amounts of informativeness and posterior probability, where larger posterior probability results in graphs that are more trimmed (less informative) and vice versa. To be more concrete, let
$$
p_{j=k} = \Pr( d_{1j} = d_{1k} | {\rm Data}), \, p_{j<k} = \Pr( d_{1j} < d_{1k} | {\rm Data}), \, p_{j>k} = \Pr( d_{1j} > d_{1k} | {\rm Data}),
$$
and define
$$
\tilde p(j,k) = \max\{p_{j=k},\, p_{j<k},\, p_{j>k}\}
$$
along with
$$
e(j,k | \tilde p) = \left\{
\begin{array}{cl}
   \{(j,k),(k,j)\}  &  \mbox{if } p_{j=k} = \tilde p(j,k); \\
   \{(j,k)\}  &  \mbox{if } p_{j<k} = \tilde p(j,k); \\
   \{(k,j)\}  &  \mbox{if } p_{j>k} = \tilde p(j,k).
\end{array}
\right.
$$
Here, $p(j,k)$ and $e(j,k)$ are computed under the assumption that $p_{j=k}$, $p_{j<k}$, and $p_{j>k}$ are all different. Under this assumption, $e(j,k)$ is uniquely determined. Based on $e(j,k)$, we now define an initial graph $\tilde E$ as
$$
\tilde E = \bigcup_{j<k} e(j,k | \tilde p).
$$

Before trimming $\tilde E$, we must verify that it represents a coherent set of treatment statements; that is, there is a transitive relation among the $K$ treatment effects. For example, if treatments~$j$ and $k$ have the same effect and so do treatments~$j$ and $l$, then treatments~$k$ and $l$ also have the same effect; or if the treatment effect of $j$ is less than that of $k$ and treatments~$j$ and $l$ have the same effect, then treatment~$k$ has a smaller effect than treatment~$l$. To verify the coherence of $\tilde E$, a sufficient condition is to check if ${\rm Pr}(\mathcal{E}=\tilde E|{\rm Data})>0$. This follows from the fact that the posterior distribution of $\mathcal{E}$ is supported on the space of all coherent graphs. If ${\rm Pr}(\mathcal{E}=\tilde E|{\rm Data})>0$ cannot be verified, we use the closest coherent graph to $\tilde{E}$, denoted as $E_0$, as the initial graph. Thus, the starting graph $E_0$ is defined as
$$
E_{0} = \min_{ \{E : \Pr(\mathcal{E}=E|{\rm Data})>0 \}} d(\tilde{E},E)
$$
with
\begin{eqnarray*}
d(\tilde{E},E) & = & \sum_{j<k} \left\{ p_{j=k}
\mathbb{I}_{\tilde E}(j,k)\mathbb{I}_{\tilde E}(k,j) 
\mathbb{I}_{ E}(j,k)\mathbb{I}_{E}(k,j) \right.\\
&  & \hspace{5mm} + \,
p_{j<k}
\mathbb{I}_{\tilde E}(j,k)(1-\mathbb{I}_{\tilde E}(k,j))
[\mathbb{I}_{E}(j,k)\mathbb{I}_{E}(k,j) + 2(1-\mathbb{I}_{E}(j,k))\mathbb{I}_{E}(k,j)]   \\
&  & \hspace{5mm} + \,
\left. p_{j>k}
(1-\mathbb{I}_{\tilde E}(j,k))\mathbb{I}_{\tilde E}(k,j)
[\mathbb{I}_{E}(j,k)\mathbb{I}_{E}(k,j) + 2\mathbb{I}_{E}(j,k)(1-\mathbb{I}_{E}(k,j))] \right\}, 
\end{eqnarray*}
where $\mathbb{I}_{E}(j,k) = 1$ if $(j,k) \in E$, and its equals 0 otherwise. Notice that $d(\cdot,\cdot)$ compares all edges in $\tilde{E}$ with those in $E$. If edges associated with treatment $j$ and $k$ have the same direction in $\tilde{E}$ and $E$ or are both bi-directional, then a value of 0 is assigned. If one edge is bi-directional and the other one is uni-directional, then a value of 1 is assigned to this comparison. If edges for $j$ and $k$ are uni-directional but with opposite directions, then a value of 2 is assigned to this comparison. The distance $d(\cdot,\cdot)$ is defined as a weighted sum of the values obtained for these comparisons, where $\tilde p(j,k)$ ($p_{j=k}$, $p_{j<k}$, or $p_{j>k}$) are used as weights. Using weights ensures that edges with high posterior probability are mostly retained.

As noted, our goal is to identify a collection of graphs $\{E_\gamma: \gamma \in [0,1]\}$ such that $p_\gamma = \Pr(\mathcal{E}=E_\gamma|{\rm Data})$ is an increasing function of $\gamma$. The idea is to define $E_\gamma \subseteq E_0$ as a graph comprised of a collection of pairwise comparisons (edges) that meet a certain probability threshold $\gamma$. We highlight the role that $\gamma$ plays in our formation of $E_\gamma$. As $\gamma \rightarrow 1$, $E_\gamma$ becomes more sparse (less informative); while as $\gamma \rightarrow 0$, $E_\gamma$ becomes $E_0$ (more informative). Our approach is to consider a sequence of thresholds (lower bounds) $\gamma \in [0,1]$ for the pairwise probabilities
$$
E_\gamma = \bigcup_{(j,k): j<k, \, p(j,k) > \gamma} e(j,k | p),
$$
where
$$
p(i,j) = 
p_{j=k} 
\mathbb{I}_{E_0}(j,k)\mathbb{I}_{E_0}(k,j) 
+
p_{j<k}
\mathbb{I}_{E_0}(j,k)(1-\mathbb{I}_{E_0}(k,j))
+
p_{j>k}
(1-\mathbb{I}_{E_0}(j,k))\mathbb{I}_{E_0}(k,j).
$$
Users will then have access to $(E_\gamma, p_\gamma)$ for an increasing sequence of $\gamma_1,\ldots, \gamma_L$ such that $E_{\gamma_{l}} \neq E_{\gamma_{l'}}$ for every $l \neq l'$. Users can then balance the posterior probability of graph vs. the sparsity of the graph for their particular setting.

With the Gaussian effects model described in Equations~\eqref{likelihood}--\eqref{lastprior}, it is not possible to formally test that two treatments have the same effect. As a result, using our approach to summarize treatment effect orderings would necessarily result in a graph that does not contain any bi-directional arrows. This motivates extending the Gaussian effects model so that ties between treatments are possible, which will be described in the next section.

%{\color{blue} AFB: Should we consider other strategies? For example, start with the mode graph and then sequentially remove the edges with the lowest posterior probability.}

\section{Extensions to the Gaussian Effects Model}
\label{sec:newmodels}

In this section, we provide details of an extension to the Gaussian effects model described in Section~\ref{sec.models}. The extension permits relaxing constraints highlighted in the previous section by allowing ties. In Section~\ref{DP.gaussian} we first detail a standard BNP approach that clusters $d_{1k}$, $2, \ldots, K$. Since the standard BNP approach does not allow clustering treatment $1$ (reference) along with the other treatments, in Section~\ref{DP.spike_slab}, we combine the standard BNP approach with spike and slab priors that allows clustering all treatments. Details associated with computation are provided in the online supplementary material. %All computation is carried out using the {\tt CBnetworkMA} package that is found in {\tt R}. 

\subsection{Dirichlet Process Gaussian Effects Model}
\label{DP.gaussian}

If ties are permitted, then $\Pr(d_{1k} = d_{1k'}) > 0$. One way to incorporate this is to use a BNP method that induces a prior on the partitions of $\mathcal{T}$\@. BNP methods for partition modeling have exploded in popularity during the past decades. For sake of conciseness, we do not provide here a detailed review, but point interested readers to several book treatises, including Chapter~23 in \cite{gelman2014bayesian}, \cite{BNPbook:2015}, and \cite{ghosal_van_der_vaart_2017}. %We are most interested in the clustering capability of BNP methods. These are a consequence of inducing a prior distribution on the space of all partitions of treatment effects in $\mathcal{T}$. This is done by modeling $d_{1k}$ with discrete random probability measure $\mathcal{P}$. Therefore, 
Our extension to the Gaussian effects model is to relax the assumption that $d_{1k} \sim N(m_d, s_d)$; we assume that $d_{1k}|\mathcal{P} \sim \mathcal{P}$ for some unknown density $\mathcal{P}$, and then assign a BNP prior to $\mathcal{P}$\@. The BNP prior we employ is the commonly used Dirichlet process (DP)\@. Thus, we change Equation~\eqref{lastprior} to the following 
\begin{align*}
 d_{1k}|\mathcal{P} & \stackrel{\text{iid}}{\sim} \mathcal{P}, \ k = 2, \ldots, K, \ \mbox{with $d_{11} = 0$}, \\
 \mathcal{P} & \sim DP(\alpha, P_0), \\ 
  P_0 & = \mbox{N}(m_d,s_d).
\end{align*} 
Here, $DP(\alpha, P_0)$ denotes a DP with concentration parameter $\alpha$ and base measure $P_0$. We refer to Equations~\eqref{likelihood}--\eqref{tau_prior} plus the prior described above as the ``DP Gaussian effects'' model.

The property of the DP that is of most interest to our current work is its almost surely discrete nature. A consequence of this property is that there is positive probability for ties among the $d_{1k}$. This is more easily seen by invoking the stick-breaking construction of a DP random probability measure \citep{Sethuraman:1994}, which dictates that $\mathcal{P}$ can be expressed as the following weighted sum of point masses $\mathcal{P}(\cdot) = \sum_{h=1}^{\infty}\pi_h \delta_{d^{\star}_h}(\cdot)$. Here, $\pi_1, \pi_2, \ldots$ denote probability weights such that $\pi_h = V_h\prod_{\ell < h}(1-V_{\ell})$ with $V_{h} \sim \mbox{Beta}(1, \alpha)$, and $\delta_x(\cdot)$ is the Dirac measure at $x$ with atoms $d^{\star}_h \sim P_0$. It is common, for computational reasons, to truncate the infinite mixture by some upper bound, say $H$ \citep{ishwaran&james:2001}. Setting $V_H = 1$ ensures that the weights still add to 1. In addition, it is common to introduce latent cluster labels $c_k$ such that $c_k = h$ implies that $d_{1k} = d^{\star}_h$ (or alternatively, $d_{1k} = d^{\star}_{c_k}$). Then, we can re-express the model for $d_{1k}$ in terms of the latent cluster labels as follows
    \begin{align*}
        d_{1k} | c_k & \sim \delta_{d^{\star}_{c_k}}, \ k = 2, \ldots, K, \mbox{with $d_{11} = 0$}; \\
%        \Pr(c_k = h|\pi_1, \ldots, \pi_H) & = \pi_h, \ h = 1, \ldots, H; \\
        c_k |\pi_1, \ldots, \pi_H & \stackrel{\text{iid}}{\sim} \sum_{h=1}^H\pi_h\delta_h, \ k = 2, \ldots, K;
        \\
        d^{\star}_h & \stackrel{\text{iid}}{\sim} P_0, \ h = 1, \ldots, H, \ P_0 = \mbox{N}(m_d, s_d).
    \end{align*}
In this form, it is straightforward to see how our extension permits treatment effects to be equal with positive probability. Since $\Pr(c_k = c_{k'}) \ne 0$, we have that $\Pr(d^{\star}_{c_k} = d^{\star}_{c_{k'}}) = \Pr(d_{1k} = d_{1k'}) \ne 0$. 

\subsection{Dirichlet Process Spike-Slab Model}
\label{DP.spike_slab}

The appeal of the DP Gaussian effects model in our setting is that $\Pr(d_{1k} = d_{1k'}) > 0$; as a result, we can consider ties. However, it is possible in an NMA that a subset of treatments have no effect, i.e., $d_{1k} = 0$ for some $k \in \{2, \ldots, K\}$. Now, due to the Gaussian base measure employed in the DP Gaussian effects model, we have that $\Pr(d_{1k} = 0) = 0$. We next describe an extension to the DP Gaussian effects model that relaxes this constraint. 

In order for $\Pr(d_{1k} = 0) > 0$, it is necessary that $\Pr(d^{\star}_h = 0) > 0$. Our approach to accommodating this characteristic is by using a spike and slab as the base measure of a DP\@. The spike and slab prior, which is commonly used in Bayesian variable selection \citep{mithcell&beauchamp:1988, george&mcculloch:1993, ishwaran&rao:2005}, is a two-component mixture model. The slab component is a relatively diffuse distribution typically centered at 0. The spike component is typically specified in one of two ways \citep{walli&wagner:2011}. One is to set the spike to a degenerate distribution at 0, while the other is to treat the spike as an absolutely continuous distribution centered at 0 with a small variance. In our formulation, we will use the latter. The reasons for doing so are two-folded. First, as mentioned by \cite{canale:2017}, some BNP methods become more complex if the base measure has an atomic component. Second, and perhaps more importantly, using a continuous distribution with a small variance will give practitioners the ability to define a zero effect within the context of their study. That is, by setting the spike component to $\mbox{N}(0, v_0/3)$, $v_0$ is such that, if $d^{\star}_{h} \in (-v_0,v_0)$, then all treatments $k$ such that $c_k = h$ are assumed to have the same mean effect as treatment 1.

Using the spike and slab in BNP modeling has a presence in the literature \citep{kim&dahl&vannucci:2009, scarpa&dunson:2009, canale:2017}. What sets us apart is our desire to have a spike with zero density for values in $(-\infty, -v_0)\cup(v_0,\infty)$ and slab with zero density for values in $(-v_0,v_0)$. To this end, for the slab component, we consider the inverse moment nonlocal prior density detailed in \cite{johnson:2010}. This density has the ability to put negligible mass on the neighborhood of a particular value (which for us is 0). For the sake of completeness, we list the density function here.
\[
 \mbox{NLP}(d^{\star}_{h} | p, r, u) = \frac{pr^{u/2}}{\Gamma(u/2p)}\{(d^{\star}_{h} - 0)^2\}^{-(u+1)/2} \exp\left[-\left\{\frac{(d^{\star}_{h} - 0)^2}{r}\right\}^{-p}\right].
\]
We set $r=1$ and $u=1$ and select $p$ based on the user supplied $v_0$. The idea is to select $p$ so that $\mbox{N}(0, v_0/3)$ and $\mbox{NLP}(p,1,1)$ are 0 on $(-\infty, -v_0)\cup(v_0,\infty)$ and $(-v_0,v_0)$, respectively. To do so, we find $x_0$ such that $\int_{-x_0}^{x_0}\mbox{N}(x | 0, v_0/3)dx \approx 0.999$ and then set $p>0$ to the value that produces $\mbox{NLP}(x_0 | p, 1, 1) = \mbox{N}(x_0 | 0, v_0/3)$. An example of the the spike and slab densities we employ when $v_0 = 0.5$ is provided in the right plot of Figure~\ref{fig:Illust_Network_Treatment}. Notice that the overlap of density between the spike and slab is essentially 0.

% \begin{figure}[hbt]
%   \begin{center}
%   \includegraphics[width=.5\textwidth]{nlp_plot.pdf}
%  \end{center}
% \caption{Spike and slab density where the spike is $N(0, v_0/3)$ and the slab is $\mbox{NLP}(p, 1, 1)$. The $p$ is selected so that $\mbox{NLP}(x_0; p, 1, 1) = \mbox{N}(x_0 | 0, v_0/3)$ where $x_0$ such that $\int_{-x_0}^{x_0}\mbox{N}(x | 0, v_0/3)dx \approx 0.999$. }
%  \label{fig:spike&slab}
% \end{figure}

With the foregoing details of the spike and slab base measure, the extension to the DP Gaussian effects model (which we call the DP Spike-Slab model) is the following:
\begin{align*}
d_{1k}|\mathcal{P}  & \stackrel{\text{iid}}{\sim} \mathcal{P},  \ k = 2, \ldots, K \ \mbox{with $d_{11} = 0$}; \\
\mathcal{P} & \sim DP(\alpha, P_0); \\ 
%P_0 & = \omega_0\delta_0(\cdot) + (1-\omega_0)NLP(k,s,v),\\ 
P_0 & = \omega_0\mbox{N}(0, v_0/3) + (1-\omega_0)\mbox{NLP}(p,1,1);\\ 
\omega_0 & \sim \mbox{Beta}(a_{\omega}, b_{\omega}),
\end{align*}
where $\omega_0$ is associated with $\Pr(d^{\star}_h = 0)$, and $a_{\omega}$ and $b_{\omega}$ are user supplied. As with the DP Gaussian effects model, introducing latent component labels for the DP ($c_k$) and for the spike and slab ($s_h$) facilitate computation. After introducing both component labels, the DP Spike-Slab model can be re-expressed as
\begin{align*}
  d_{1k} | c_k & \stackrel{\text{ind}}{\sim} \delta_{d^{\star}_{c_k}}, \ k = 2, \ldots, K, \mbox{with $d_{11} = 0$}; \\
%        \Pr(c_k = h|\pi_1, \ldots, \pi_H) & = \pi_h, \ h = 1, \ldots, H; \\
        c_k |\pi_1, \ldots, \pi_H & \stackrel{\text{iid}}{\sim} \sum_{h=1}^H\pi_h\delta_h, \ k = 2, \ldots, K;  \\
  d^{\star}_h|s_h & \sim \mbox{N}(0, v_0/3)^{s_h}~\mbox{NLP}(p, 1, 1)^{1-s_h},  \ h = 1, \ldots, H;\\
    s_h & \sim \mbox{Bernoulli}(\omega_0), \ h = 1, \ldots, H;\\
    \omega_0 & \sim \mbox{Beta}(a_{\omega}, b_{\omega}),
\end{align*}
where, once again, $p$ is determined by $v_0$.

\section{Simulation Study}
\label{sec:sim}

\subsection{Simulation Designs}
\label{sec:simdesign}

We conducted a simulation study to illustrate our approach and highlight its utility when there exist treatment effects that are similar and/or close to 0. We also explored the loss of efficiency if no such treatment effects exist. The treatment network design is based on that found in the antidepressant dataset described in Section~\ref{sec:data}. However, to make the simulation study more manageable, we only consider 6 of the 12 treatments, and this subset of treatments appear in $i = 1, \ldots, 55$ studies. Data were generated by fixing $d_{1k}$ for $k \in \{1, \ldots, 6\}$ to specific values, which were then used to generate $\delta_{i,b_ik}$ that in turn produced values for $p_{ik}$. The $p_{ik}$ values were then used to generate the number of successes (i.e., $y_{ik}$) based on the number of trials (i.e., $n_{ik}$). The $n_{ik}$ values all came from the subset of the antidepressant dataset described in Section~\ref{sec:data}.

In the simulation, we studied the impact of three factors: (i) the signal-to-noise ratio; (ii) the effect/treatment structure; and (iii) the effect size. The signal vs.\ noise was controlled by $\tau$, and we considered $\tau \in \{0.05, 0.1\}$. The effect/treatment structure was determined by the values of $\bm{d}_1$. In order to select reasonable values for $\bm{d}_1$, we fit the DP Spike-Slab model to the subset of the antidepressant dataset. The posterior mean of $\bm{d}_1$ from this fit was $(0, 0.3, 0.0, 0.0, 0.3, 0.3)'$, which is the basis of the three different $\bm{d}_1$ vectors that we employed. Each of the three $\bm{d}_1$ vectors was constructed to favor one of the three models under consideration (the Gaussian effects, DP Gaussian effects, or DP Spick-Slab models). Their exact forms are provided in the following enumerated list. 
\begin{enumerate}
    \item $\bm{d}_1 = (0, -0.6, -0.3, 0.3, 0.6, 0.9)'$. This $\bm{d}_1$ vector favors the Gaussian effects model as all entries are different and none are equal to 0. We refer to this data generating scenario as ``$\bm{d}_1 \sim$ Gaussian.''
    \item $\bm{d}_1 = (0, 0.3, -0.3, -0.3, 0.3, 0.3)'$. This $\bm{d}_1$ vector favors the DP Gaussian effects model as there are two clusters of effects and none are equal to 0. We refer to this data generating scenario as ``$\bm{d}_1 \sim$ DP Gaussian.''
    \item $\bm{d}_1 = (0, 0.3, 0.0, 0.0, 0.3, 0.3)'$. This $\bm{d}_1$ vector favors the DP Spike-Slab model as there are two clusters with one setting the effects to 0. We refer to this data generating scenario as ``$\bm{d}_1 \sim$ DP Spike-Slab.''
\end{enumerate}
Finally, to study the effect size, we multiplied each of the $\bm{d}_1$ vectors in the three scenarios above by 0 (which resulted in no nonzero effects), 1, and 2 (which corresponded to an effect size that was twice as large). For simplicity, these three ``effect sizes'' are called effect size 0, effect size 1, and effect size 2, respectively. In total, there were 18 data generating scenarios; under each scenario, 100 datasets were generated.

To each generated data set, we fit the Gaussian effects, DP Gaussian effects, and DP Spike-Slab models. For all three models, we set $m_b = 0$, $s_b=10$, $m_{\ell} = -2.34$, and $s_{\ell} = 2$. The later two prior values were selected based on suggestions from \cite{turner_etal:2012}. For the Gaussian effects and DP Gaussian effects models, we set $m_d = 0$ and $s_d = 1$. For the DP Gaussian effects and DP Spike-Slab models, we set $\alpha = 1$. For the DP Spike-Slab model, we set $a_{\omega} = b_{\omega} = 1$, and for the nonlocal prior component, we employed $v_0=0.05$. We also considered $v_0 = 0.1$ with results being almost identical. All model fits were carried out by running 5 independent MCMC chains each for 200,000 MCMC samples, of which the first 100,000 were discarded as burn-in and the remaining samples were thinned by 100. We were very liberal in the amount of burn-in and thinning to ensure that convergence was met and that independent (at least approximately) samples were collected. Combining the 5 chains resulted in a total of 5,000 MCMC samples.

\subsection{Simulation Results}
\label{sec:simrslt}

We compared the three methods' performance by calculating posterior probabilities associated with the pairwise comparison graph determined by $\bm{d}_1$. We began by calculating the joint posterior probability of the true pairwise comparison graph in its entirety (i.e., all 15 comparisons jointly). Results are provided in Figure~S.1 of the online supplementary material. Notice that when ``$\bm{d}_1 \sim$ Gaussian'' was used to generate data and the effect size was 1 (i.e., the true vector $\bm{d}_1$ was multiplied by 1), the Gaussian effects model had a high probability of recovering the graph, while the DP Gaussian effects and DP Spike-Slab had much lower probabilities. This was expected as the data generating scenario favored the Gaussian effects model. However, as the effect size grew, the DP procedures performed on par with the Gaussian effects model, even when the Gaussian effects model had an advantage. When ``$\bm{d}_1 \sim$ DP Gaussian'' was used to generate data, by construction, the Gaussian effects model had no hope of recovering the graph. As expected, the DP Gaussian effects model performed best in this scenario with an effect size of 1. However, somewhat surprisingly, the DP Spike-Slab method outperformed the DP Gaussian effects when the effect size was 2 even when the DP Gaussian effects had an inherent advantage. When ``$\bm{d}_1 \sim$ DP Spike-Slab'' was used, the Gaussian effects and DP Gaussian effects procedures were not able to recover the true graph with any positive probability. These trends held even as the signal-to-noise ratio increased. If the effect size was null, then the DP Spike-Slab was the only procedure that was able to recover the joint graph with positive probability. The upshot to Figure~\ref{fig:graph_prob} is that the DP Spike-Slab procedure was the only procedure with enough flexibility to perform reasonably well in all scenarios.

% \begin{figure}[hbt]
%   \begin{center}
%   \includegraphics[width=.7\textwidth]{graphProb.pdf}
%  \end{center}
% \caption{Posterior probability associated with the entire pairwise comparison graph that was used to generate data. This graph is comprised of 15 pairwise comparisons; as a result, the probabilities displayed are joint probabilities in the sense that they are associated with the probability of 15 pairwise comparisons simultaneously occurring.}
%  \label{fig:graph_prob}
% \end{figure}

\begin{figure}[hbt]
  \begin{center}
    \includegraphics[scale = 0.48, trim={0cm 0cm 5.9cm 0cm}, clip]{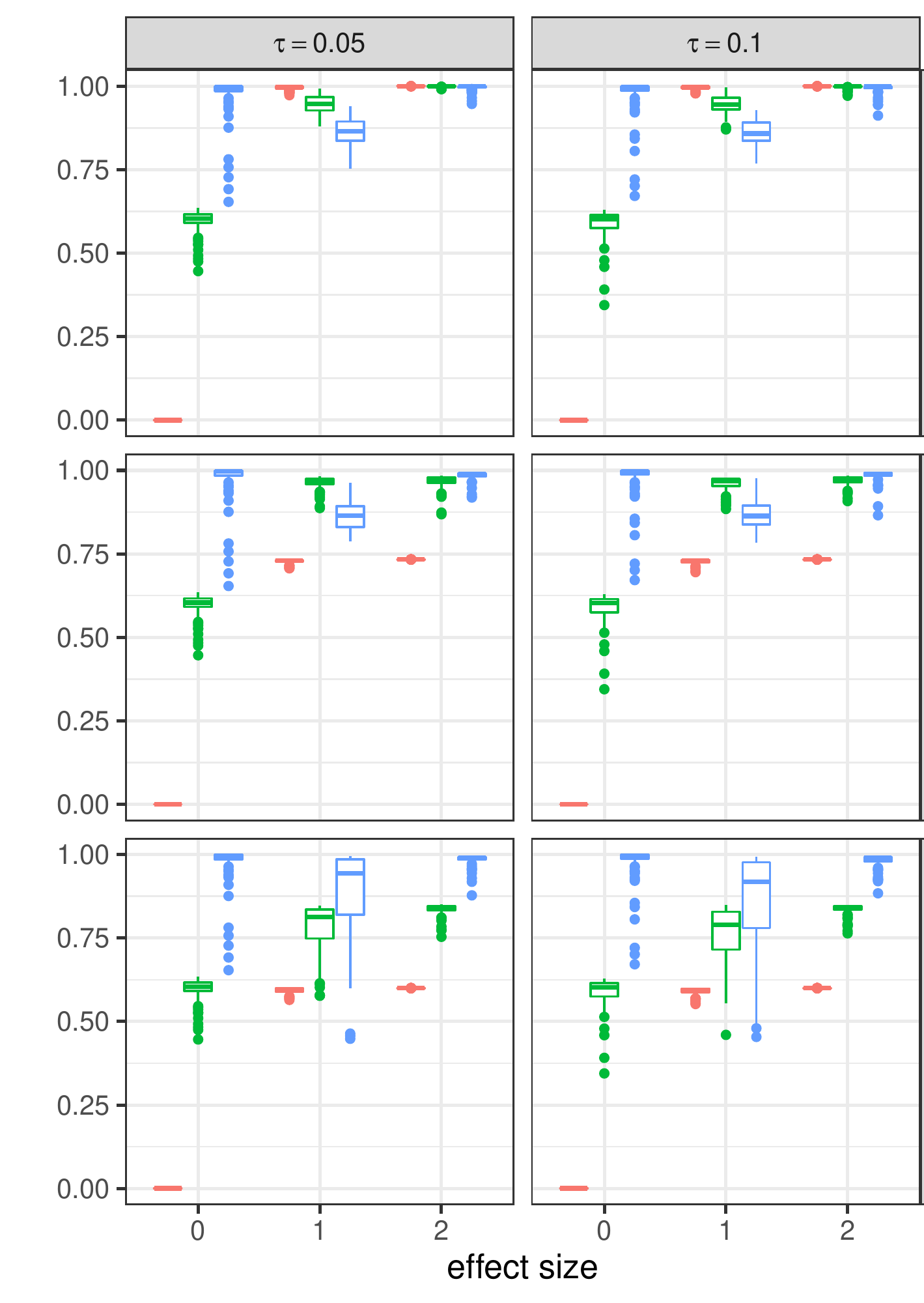}
   \includegraphics[scale = 0.48, trim={1.75cm 0cm 0cm 0cm}, clip]{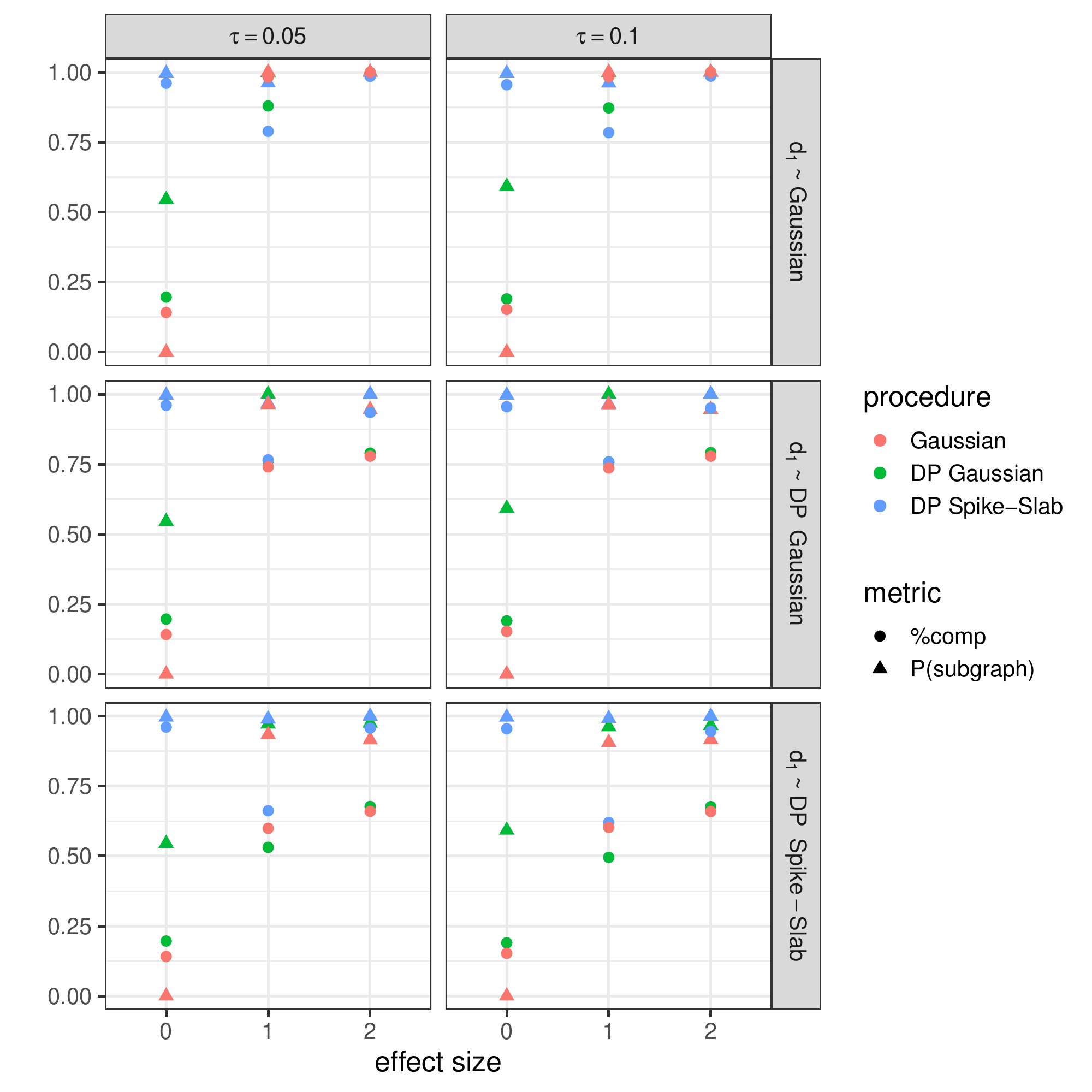}
 \end{center}
\caption{
The left two columns display the average posterior probability of true pairwise treatment comparisons. The right two columns display results from applying the method described in Section~\ref{sec.ordering.statements} to summarize 15 treatments in the simulation study. The triangle symbol corresponds to the average posterior probability associated with the estimated sub-graph. Circles correspond to the sparsity of the estimated sub-graph. The lower the value, the more sparse and the smaller number of comparisons.} 
 \label{fig:post_hoc}\label{fig:pairwise_prob}
\end{figure}

At first glance, it may seem concerning that the joint posterior probability of the graph based on the DP Spike-Slab was relatively small when generating data using ``$\bm{d}_1 \sim$ Gaussian'' and the effect size was 1. To provide additional insights into why this occurred, we also calculated the posterior probabilities for each of the individual pairwise comparisons separately. These results are provided in the left two columns of Figure~\ref{fig:pairwise_prob}. Even in the scenarios in which the DP Spike-Slab scenario had a low posterior probability of recovering the entire graph, it missed the complete graph by only one or two pairwise comparisons. This was particularly true when the effect size was doubled or null. Therefore, it seems that although the DP Spike-Slab approach may not recover the entire pairwise comparison graph with high probability in some scenarios, it only misses by one or two comparisons.

% \begin{figure}[hbt]
%   \begin{center}
%   \includegraphics[width=.7\textwidth]{pairwiseProb.pdf}
%  \end{center}
% \caption{The posterior probability of true pairwise treatment comparisons averaged over the 15 pairwise comparisons.}
%  \label{fig:pairwise_prob}
% \end{figure}

To further explore the similarity between graphs produced by the three methods that have high posterior probability and the true data generating graph, we applied the proposed method of producing treatment rankings described in Section~\ref{sec.ordering.statements}. In particular, we identified the sub-graph with the largest number of connected nodes that had a posterior probability of at least 0.9. We then computed the probability that the sub-graph identified by our approach equaled the corresponding true sub-graph. In addition, to get a sense of the denseness of the sub-graphs, we also found the percent of comparisons that were included in the estimated sub-graph (out of 15). This was done for each of the three procedures, and the results are provided in the right two columns of Figure~\ref{fig:post_hoc}. In these plots, the posterior probability of the estimated sub-graph equaling the true sub-graph is provided in addition to the percent of comparisons included in the estimated sub-graph all averaged over the 100 datasets. The first thing to notice is that, in all scenarios, the DP Spike-Slab procedure was as good as or better than the other two procedures in finding a sub-graph that had a high posterior probability of being correct. In addition, the sub-graphs estimated using the DP Spike-Slab procedure were always denser compared to those coming from the other two procedures. The upshot of this exercise is that the DP Spike-Slab procedure was able to identify dense sub-graphs with a high posterior probability of being correct in all data generating scenarios, while the Gaussian effects method only performed well in scenarios that favored that particular method.

\section{Analysis of the Antidepressants Dataset}
\label{sec:realdata}

We now proceed to apply the DP Spike-Slab approach to the data described in Section~\ref{sec:data}. The Gaussian effects model is also fit to provide context. Each of the two models were fit by collecting 3000 total MCMC samples from three independent chains with disparate starting values. Each chain was thinned by 100 iterates after discarding the first 100,000 iterates as burn-in. For both methods we set $m_b = 0$, $s_b=10$, $m_{\ell} = -2.34$, and $s_{\ell} = 2$. The later two hyperprior values are recommended in \cite{turner_etal:2012}. For the DP Spike-Slab model we also set $a_{\omega} = b_{\omega} = 1$, $\alpha = 1$, and $v_0=0.1$. 

For the DP Spike-Slab approach, $d_{1k}$ can be exactly equal to 0 or $d_{1k'}$. For this reason, achieving an credible interval of arbitrary posterior probability level might not be possible, and the construction of credible intervals requires special attention (see Section~4.3 of \citealt{womack:2014}). We summarize the posterior distribution of treatment effects using $p_{j=k}$, $p_{j>k}$, and $p_{j<k}$, which were introduced in Section~\ref{sec.ordering.statements}. In particular, we use the following criteria that produces a conditional type of a $(1-\alpha)$ credible interval. Let $o_{kk'} = \exp(d_{1k'} - d_{1k})$ denote the odds ratio of treatment~$k$ compared to treatment~$k'$. Then, using a strategy that resembles that in \cite{womack:2014}, we define a $(1-\alpha)$-conditional credible interval as
\begin{align*}
{\rm CI}(o_{jk}) = \left\{
\begin{array}{cl}
    \{1\} & \mbox{if } p_{j=k} = p(j,k); \\
    \left[{\rm quantile}_{1-(1-\alpha/2)p_{j>k}}(o_{jk}) , {\rm quantile}_{1-(\alpha/2)p_{j>k}}(o_{jk}) \right] & \mbox{if } p_{j>k} = p(i,j); \\
    \left[{\rm quantile}_{(\alpha/2)p_{j<k}}(o_{jk}) , {\rm quantile}_{(1-\alpha/2)p_{j<k}}(o_{jk}) \right] & \mbox{if } p_{j<k} = p(i,j).
\end{array}
\right.
\end{align*}
We then report ${\rm CI}(o_{jk})$ along with $\Pr( o_{jk} \in {\rm CI}(o_{jk}) | {\rm Data})$, where
$$\Pr( o_{jk} < {\rm quantile}_{a}(o_{jk}) | {\rm Data}) = a.$$
Notice that if $p_{j>k} = 1$ or $p_{j<k} = 1$ then $\Pr( o_{jk} \in {\rm CI}(o_{jk}) | {\rm Data}) = (1-\alpha)$.
%}

\begin{figure}[hbt]
   \centering
   \begin{subfigure}{0.4\textwidth}
   \centering
   \caption{$\Pr(\mathcal{E} = \mbox{mode graph} |  {\rm Data}) < 0.01$}
   \label{fig:Cipriani_network_model1.1}
   \includegraphics[scale = 0.4, page = 2, width = \textwidth, trim={1.9cm 2cm 0.9cm 1.9cm}, clip]{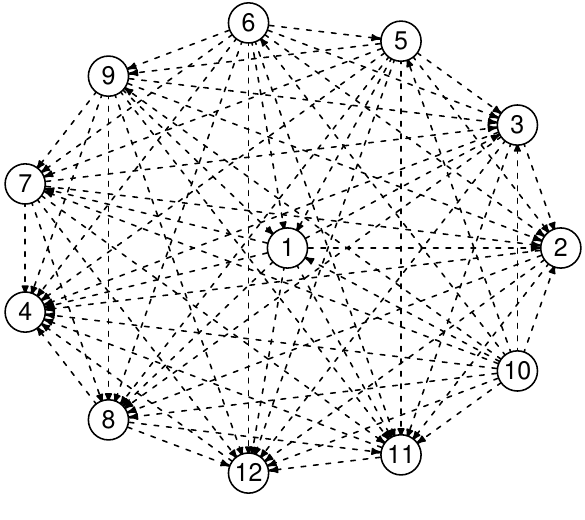}
   \end{subfigure}
   \begin{subfigure}{0.4\textwidth}
   \centering
   \caption{$\Pr(\mathcal{E} = E_{\gamma} | {\rm Data}) < 0.01$, $\gamma = 0.33$}
   \label{fig:Cipriani_network_model1.2}
   \includegraphics[scale = 0.4, page = 3, width = \textwidth, trim={1.9cm 2cm 0.9cm 1.9cm}, clip]{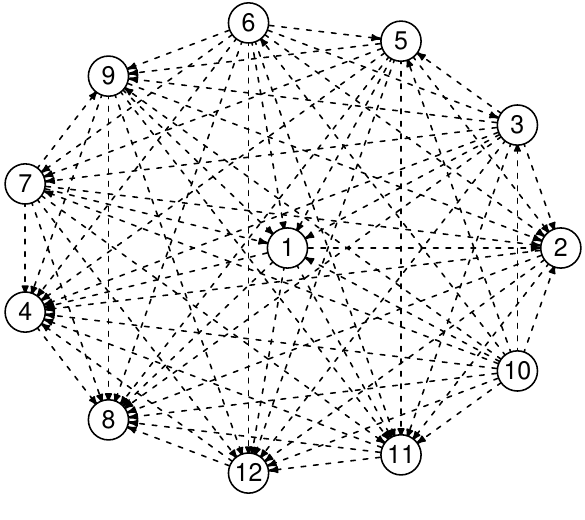}
   \end{subfigure}
   \\
   \begin{subfigure}{0.4\textwidth}
   \centering
   \caption{$\Pr(\mathcal{E} = E_{\gamma} | {\rm Data}) = 0.59$, $\gamma = 0.88$}
   \label{fig:Cipriani_network_model1.3}
   \includegraphics[scale = 0.4, page = 58, width = \textwidth, trim={1.9cm 2cm 0.9cm 1.9cm}, clip]{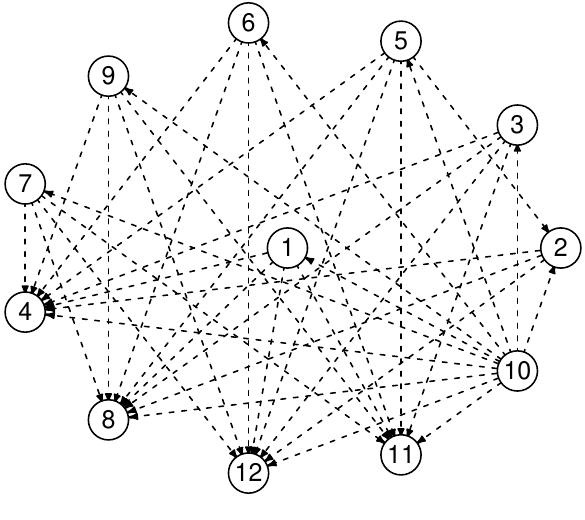}
   \end{subfigure}
   \begin{subfigure}{0.4\textwidth}
   \centering
   \caption{$\Pr(\mathcal{E} = E_{\gamma} | {\rm Data}) = 0.77$, $\gamma = 0.96$}
   \label{fig:Cipriani_network_model1.4}
   \includegraphics[scale = 0.4, page = 66, width = \textwidth, trim={1.9cm 2cm 0.9cm 1.9cm}, clip]{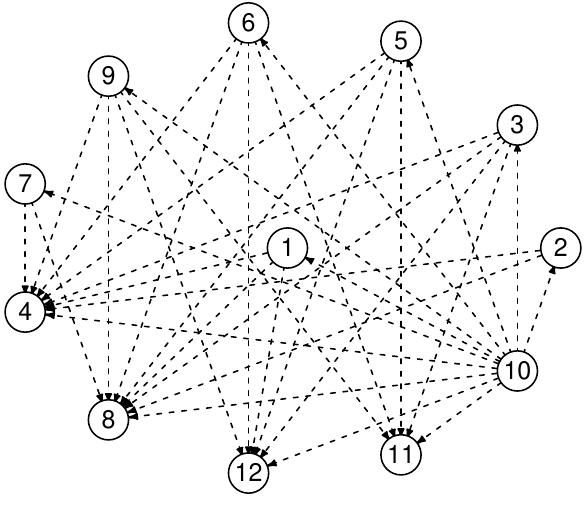}
   \end{subfigure}
   \\
   \begin{subfigure}{0.4\textwidth}
   \centering
   \caption{$\Pr(\mathcal{E} = E_{\gamma} | {\rm Data}) = 0.89$, $\gamma = 0.98$}
   \label{fig:Cipriani_network_model1.5}
   \includegraphics[scale = 0.4, page = 68, width = \textwidth, trim={1.9cm 2cm 0.9cm 1.9cm}, clip]{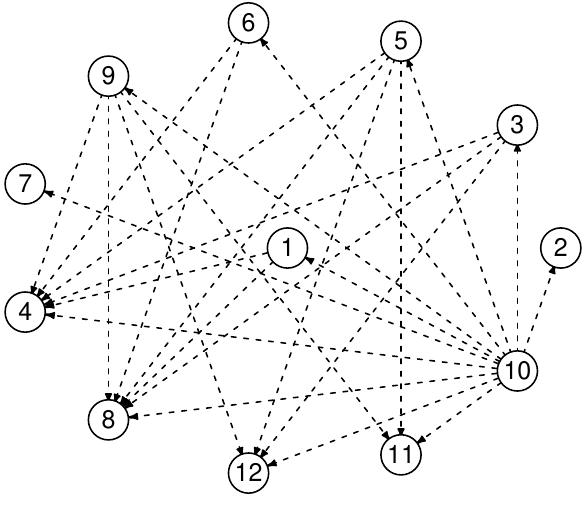}
   \end{subfigure}
   \begin{subfigure}{0.4\textwidth}
   \centering
   \caption{$\Pr(\mathcal{E} = E_{\gamma} | {\rm Data}) = 0.97$, $\gamma = 0.99$}
   \label{fig:Cipriani_network_model1.6}
   \includegraphics[scale = 0.4, page = 69, width = \textwidth, trim={1.9cm 2cm 0.9cm 1.9cm}, clip]{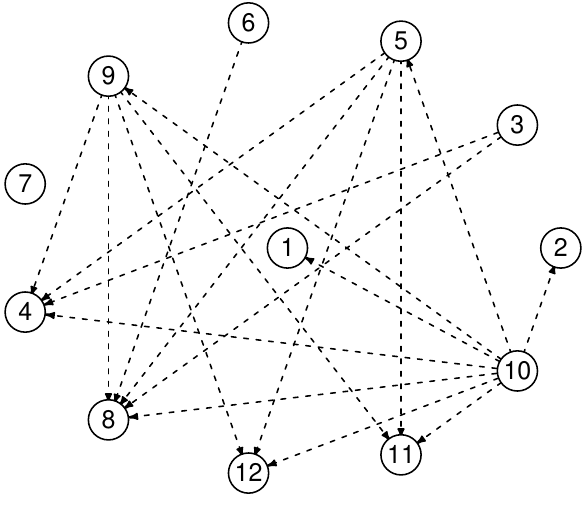}
   \end{subfigure}
   \\
\caption{Posterior network treatment effects based on the Gaussian effects model. A dashed one-directional arrow from A to B indicates that treatment~B performs better treatment~A\@. A solid bi-directional arrow between A and B indicates that treatments~A and B have the same performance.}
 \label{fig:Cipriani_network_model1.pdf}
\end{figure}

\begin{figure}[hbt]
  \begin{center}
   \begin{subfigure}{0.4\textwidth}
   \centering
   \caption{$\Pr(\mathcal{E} = \mbox{mode graph} |  {\rm Data})=0.36$}
   \label{fig:Cipriani_network_model3.1}
   \includegraphics[scale = 0.4, page = 2, width = \textwidth, trim={1.9cm 2cm 0.9cm 1.9cm}, clip]{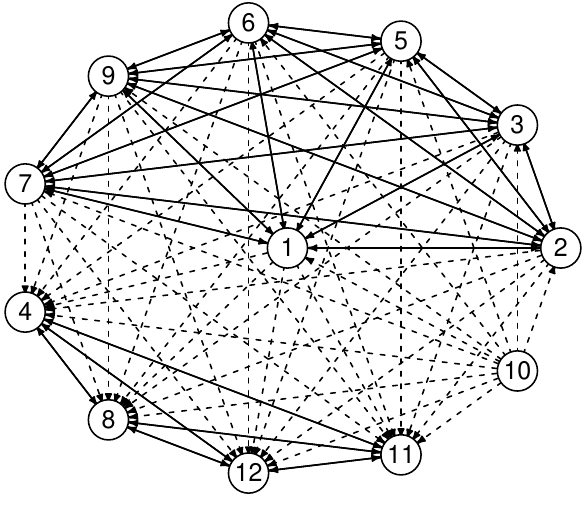}
   \end{subfigure}
   \hspace{0.05\textwidth}
   \begin{subfigure}{0.4\textwidth}
   \centering
   \caption{$\Pr(\mathcal{E} = E_{\gamma} |  {\rm Data})=0.36$, $\gamma = 0.33$}
   \label{fig:Cipriani_network_model3.2}
   \includegraphics[scale = 0.4, page = 3, width = \textwidth, trim={1.9cm 2cm 0.9cm 1.9cm}, clip]{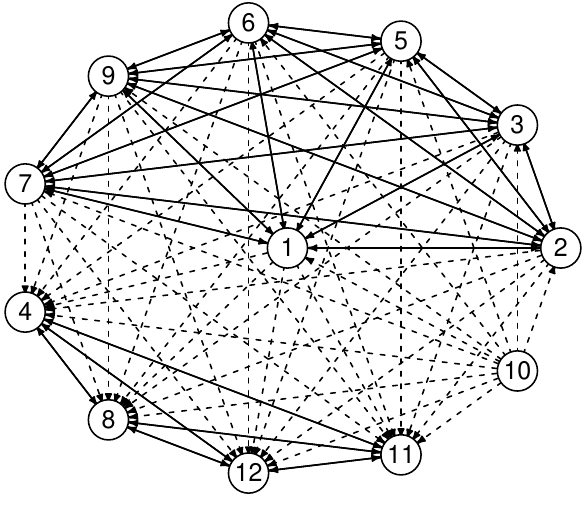}
   \end{subfigure}
   \\
   \begin{subfigure}{0.4\textwidth}
   \centering
   \caption{$\Pr(\mathcal{E} = E_{\gamma} |  {\rm Data})=0.57$, $\gamma = 0.84$}
   \label{fig:Cipriani_network_model3.3}
   \includegraphics[scale = 0.4, page = 54, width = \textwidth, trim={1.9cm 2cm 0.9cm 1.9cm}, clip]{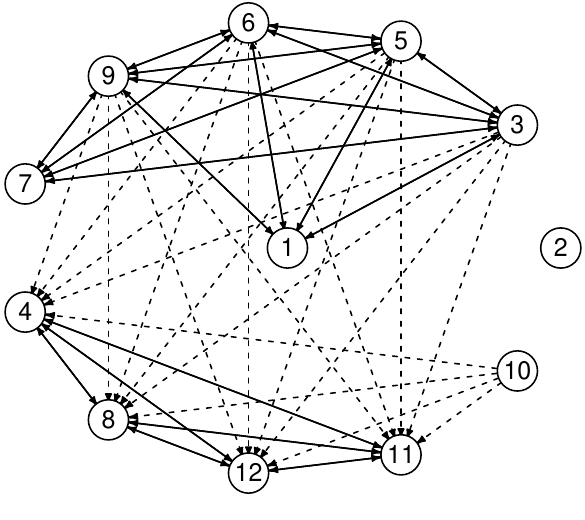}
   \end{subfigure}
   \hspace{0.05\textwidth}
   \begin{subfigure}{0.4\textwidth}
   \centering
   \caption{$\Pr(\mathcal{E} = E_{\gamma} |  {\rm Data})=0.78$, $\gamma = 0.88$}
   \label{fig:Cipriani_network_model3.4}
   \includegraphics[scale = 0.4, page = 58, width = \textwidth, trim={1.9cm 2cm 0.9cm 1.9cm}, clip]{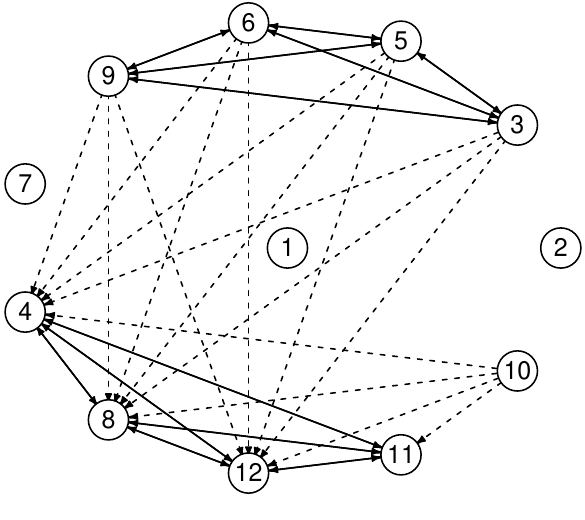}
   \end{subfigure}
   \\
   \begin{subfigure}{0.4\textwidth}
   \centering
   \caption{$\Pr(\mathcal{E} = E_{\gamma} |  {\rm Data})=0.82$, $\gamma = 0.93$}
   \label{fig:Cipriani_network_model3.5}
   \includegraphics[scale = 0.4, page = 63, width = \textwidth, trim={1.9cm 2cm 0.9cm 1.9cm}, clip]{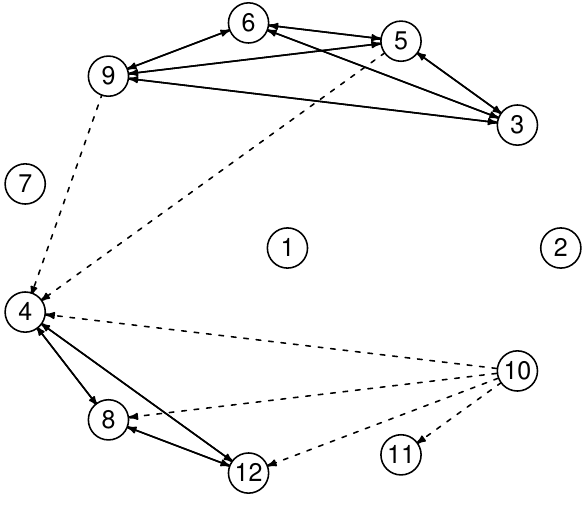}
   \end{subfigure}
   \hspace{0.05\textwidth}
   \begin{subfigure}{0.4\textwidth}
   \centering
   \caption{$\Pr(\mathcal{E} = E_{\gamma} |  {\rm Data})=0.96$, $\gamma = 0.97$}
   \label{fig:Cipriani_network_model3.6}
   \includegraphics[scale = 0.4, page = 67, width = \textwidth, trim={1.9cm 2cm 0.9cm 1.9cm}, clip]{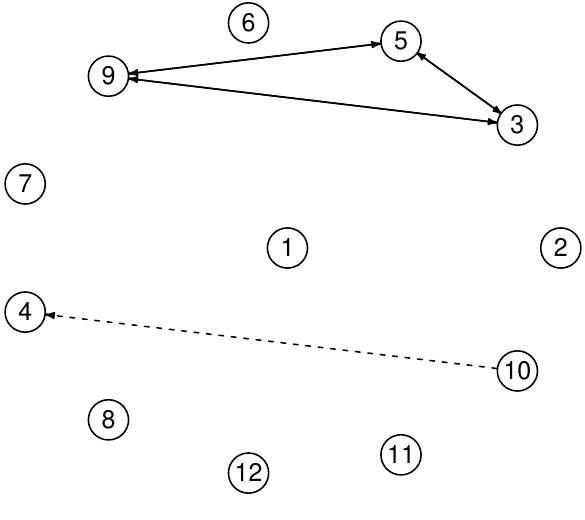}
   \end{subfigure}
 \end{center}
\caption{Posterior network treatment effects based on the DP Spike-Slab model. A dashed one-directional arrow from A to B indicates that treatment~B performs better treatment~A\@. A solid bi-directional arrow between A and B indicates that treatments~A and B have the same performance.}
 \label{fig:Cipriani_network_model3.pdf}
\end{figure}

Figures~\ref{fig:Cipriani_network_model1.pdf} and \ref{fig:Cipriani_network_model3.pdf} display the posterior network treatment effects based on the Gaussian effects and DP Spike-Slab models, respectively. The graphs' interpretations are detailed in Section~\ref{sec.ordering.statements} (Figure~\ref{fig:Illust_Network_Treatment}). The conventional Gaussian effects model was not able to identify the relationships of exactly the same treatment performance, so all treatment relationships in the graphs in Figure~\ref{fig:Cipriani_network_model1.pdf} were visualized as one-directional arrows. Based on the cumulative probabilities of being among the four most efficacious treatments, the original study of \cite{cipriani2009comparative} concluded that the best treatments were mirtazapine (24.4\%), escitalopram (23.7\%), venlafaxine (22.3\%), sertraline (20.3\%), citalopram (3.4\%), milnacipran (2.7\%), bupropion (2.0\%), duloxetine (0.9\%), fluvoxamine (0.7\%), paroxetine (0.1\%), fluoxetine (0.0\%), and reboxetine (0.0\%), i.e., treatments~8, 4, 12, 11, 2, 7, 1, 3, 6, 9, 5, and 10, accordingly. Treatments~8, 4, 12, and 11 seem to be in a cluster of best treatments because their cumulative probabilities were all about 20\% to 25\% and noticeably larger than the remaining treatments.

Our re-analyses based on the Gaussian effects model were generally consistent with these original findings, while there were a few exceptions. Specifically, in Figure~\ref{fig:Cipriani_network_model1.1}, the full relationships between all 66 treatment comparisons indicate that treatment~4 was the best, followed by treatments~12, 11, 8, 2, 3, 7, 1, 9, 5, 6, and 10, accordingly. These differences in the rankings from the original conclusions were not surprising; the cumulative probabilities for treatment rankings in the original analyses were fairly similar, and some differences could be due to Monte Carlo sampling errors when performing the Bayesian analyses. Without trimming any treatment comparisons, the foregoing exact relationships between all treatment comparisons occurred only with a posterior joint probability of $<$0.01. This further suggested that these relationships may not be reliable for practical applications.

Figures~\ref{fig:Cipriani_network_model1.2}--\ref{fig:Cipriani_network_model1.6} present the network treatment effects after trimming some uncertain treatment comparisons so that the $\tilde{p}(j, k)$'s defined in Section~\ref{sec.ordering.statements} are beyond a specific threshold $\gamma$ for all comparisons $j$ vs.\ $k$ connected with arrows. For example, if we require $\gamma = 0.96$, the number of eligible treatment comparisons was reduced to 34, and the posterior joint probability of the network relationships became 0.77 (Figure~\ref{fig:Cipriani_network_model1.4}). The cluster of best treatments, 8, 4, 12, and 11, identified in the original analyses were no longer connected with arrows, so their relationships became uncertain under the restriction of $\gamma$. If we increased $\gamma$ to 0.98, then the posterior joint probability increased to 0.89 (Figure~\ref{fig:Cipriani_network_model1.5}). If the threshold was further increased to 0.99, then the posterior joint probability became 0.97 (Figure~\ref{fig:Cipriani_network_model1.6}). Such a high joint probability could generally ensure the reliability of treatment comparisons. For example, in this case, we could conclude that treatments~8 was better than treatments~3, 5, 6, 9, and 10, treatment~4 was better than treatments~3, 5, 9, and 10, and each of treatments~11 and 12 was better than treatments~5, 9, and 10. No reliable conclusion could be drawn about treatment~7.

Using our proposed DP Spike-Slab model, we were able to identify treatments with exactly the same performance. Without trimming any uncertain treatment comparisons, Figure~\ref{fig:Cipriani_network_model3.1} shows the full network relationships. Treatments~4, 8, 11, and 12 were formally identified to have the same performance, supporting the previous observation based on the original analyses. Their performance was better than treatments~1, 2, 3, 5, 6, 7, and 9, which also had the same performance. Treatment~10 had the worst performance. However, the posterior joint probability of these network relationships was only 0.36. As in the analyses using the Gaussian effects model, we increased the threshold $\gamma$ to achieve more satisfactory joint probabilities. If the threshold was set to 0.93, the posterior joint probability became 0.82. We conclude that treatments~4, 8, and 12 had the same best treatment. Treatment~11 was no longer in this cluster; nevertheless, it was better than treatment~10. In addition, treatments~3, 5, 6, and 9 have the same performance, and they were worse than the group of treatments~4, 8, and 12. We could not draw reliable conclusions about treatments~1, 2, and 7. If the threshold was further set to 0.97, the posterior joint probability increased to 0.96. In this case, we could only reliably conclude that treatments~3, 5, and 9 had the same performance, and treatment~4 was better than 10. No reliable conclusions could be drawn for other treatment comparisons.

Table~\ref{fig:realdatarslt} gives the league table of results for all treatment comparisons based on both the Gaussian effects and DP Spike-Slab models. For the DP Spike-Slab model, the league table reports the posterior probability of treatments being equal for each comparison and the posterior probability of treatment effect belonging to the interval estimate. For the conventional Gaussian effects model, because it could not model the cases of exact equal treatment effects, the posterior probability of treatments being equal was always 0, and the posterior probability of treatment effect belonging to the interval estimate was always 95\%. The posterior mean estimates of odds ratios by the DP Spike-Slab model were generally similar to those by the Gaussian effects model, while there were still some noticeable differences. The interval estimates by the two models were substantially different for most comparisons. For example, the odds ratio of treatment~2 vs.\ 1 was estimated as 1.03 by the Gaussian effects model and as 1.04 by DP Spike-Slab; both estimates were close to the null value of 1. Nevertheless, the Gaussian effects model could not conclude that the odds ratio was exactly 1; it could only give a 95\% credible interval of [0.84, 1.26]. On the other hand, the credible interval by the DP Spike-Slab model was [1.00, 1.00], and the posterior probability of the odds ratio equal 1 was 74.57\%. For another comparison, treatment~8 vs.\ 7, the Gaussian effects model gave a posterior mean odds ratio of 1.37 with 95\% credible interval [0.99, 1.84], covering 1. DP Spike-Slab gave an estimate of 1.23, with a probability of 20.13\% that these two treatments have exactly the same effect. The interval estimate was [1.17, 1.39], and the posterior probability of treatment effect belonging to this interval was 76.53\%.

\begin{table}[t]
\centering
\includegraphics[width = \textwidth]{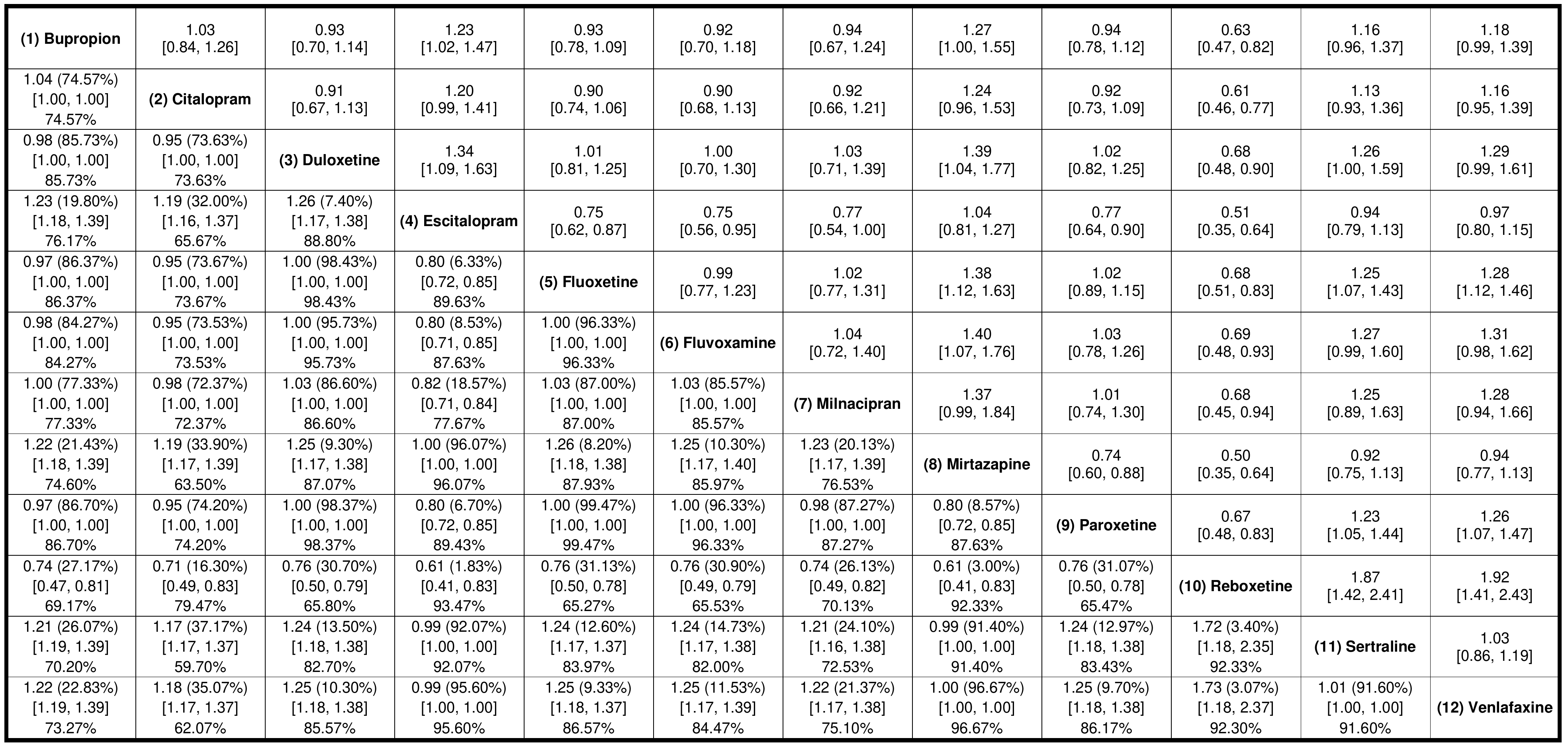}
\caption{League table of the network meta-analysis of antidepressants. The upper triangular cells are produced by the Gaussian effects model, the lower triangular cells are produced by the DP Spike-Slab model, and the diagonal cells present the treatment names. Each data cell includes the posterior mean estimate of the odds ratio (the treatment in the later alphabetical order vs.\ that in the former alphabetical order), with 95\% credible intervals in parentheses. For the DP Spike-Slab model, we report intervals estimated using method as described in Section~\ref{sec:realdata}; in each cell, we also report the posterior probability of treatments being equal in parentheses in the first line and that of treatment effect belonging to the reported interval in the last line.}
\label{fig:realdatarslt}
\end{table}

%{\color{blue} AFB: Maybe we can represent the magnitude of the effects (Posterior mean) or the pairwise probabilities with the thickness of the edge.}

\section{Discussion} \label{sec:disc}

This article has proposed a novel nonparametric Bayesian approach for NMA. The approach overcomes several limitations of existing NMA, including failure to account for uncertainties in treatment rankings. By employing a spike and slab distribution, the treatment ordering statement of our Bayesian approach can model the cases of exactly equal treatment effects (i.e., treatments with the same rank). The approach can also yield posterior network treatment effects showing the relationships among treatments and the corresponding posterior joint probability of having such relationships. This framework is entirely different from the current practice of interpreting each pair of treatment comparisons separately, which could lead to exaggerated treatment effects caused by multiplicity issues. The simulation study showed that our method is able to accurately recover treatment rankings when equal treatment effects are present in an NMA. In addition, it also showed that this added flexibility comes at a limited cost when equal treatments do not exist, and conventional NMA methods have no hope of producing accurate treatment rankings when there exist treatments whose effects are essentially the same.

Nevertheless, this paper has some limitations. First, we only considered NMAs with binary outcomes. The proposed methods could be broadly generalized to other types of outcomes (e.g., continuous or survival), and more efforts are needed to generalize them \citep{dias2013evidence}. Second, the methods are developed under the assumption of evidence consistency, i.e., direct comparisons share the same overall effects as indirect comparisons. However, this might not be valid in some cases; for example, some comparisons could be contaminated by outliers, or some effect modifiers should be incorporated in a network meta-regression \citep{higgins2012consistency, jansen2013network}. The proposed methods could be potentially extended to deal with inconsistency. For example, inconsistency is generally modeled via introducing additional parameters in an NMA model, either as treatment-loop-specific parameters or design-by-treatment interactions \citep{bucher1997results, chaimani2017common}. These inconsistency parameters may dramatically increase model complexity and thus lead to treatment effect estimates with larger variations. Nevertheless, inconsistency may actually exist in a few loops or design-by-treatment interactions for real-world data, and most inconsistency parameters are truly 0. The nonparametric Bayesian approach with the spike and slab distribution could be similarly used to permit inconsistency parameters being exactly estimated as 0. Third, the induced random partition model from a DP can be restrictive. New approaches may be developed to allow incorporating prior information on partition probabilities. For example, it might prove useful to include information regarding treatment composition and/or clinicians' expertise in the partition model.

\section*{Acknowledgements}

LL was supported in part by the US National Institutes of Health/National Institute of Mental Health grant R03~MH128727 and National Institutes of Health/National Library of Medicine grant R01~LM012982. 

\begin{spacing}{1.55}
\bibliographystyle{apalike}
\bibliography{reference}
\end{spacing}
\end{document}

\def\spacingset#1{\renewcommand{\baselinestretch}%
{#1}\small\normalsize} \spacingset{1}

%%%%%%%%%%%%%%%%%%%%%%%%%%%%%%%%%%%%%%%%%%%%%%%%%%%%%%%%%%%%%%%%%%%%%%%%%%%%%%

\maketitle

This document contains supplementary material to the paper `` Nonparametric Bayesian Approach to Treatment Ranking in Network Meta-Analysis with Application to Comparisons of Antidepressants''.

\section{Computational Details}\label{supp:sec.comp}

\subsection{Computation Details Associated with MCMC Algorithm} \label{sec:computation}

In this section, we provide a few brief details associated with computation necessary to sample from the joint posterior distribution based on the the Gaussian effects, DP Gaussian effects and the DP Spike-Slab models. The update steps for $\mu_{i, b_i}$, $\delta_{i, b_ik}$, and $\tau$ in the MCMC algorithm constructed for the three models are the same. Hence, we begin with these. 

Before doing so, we provide an alternative way of expressing equations (4) and (5) found in the main document that requires introducing two pieces of notation.  Let $t_i = |\mathcal{T}^{-b_i}_i|$ denote the number of treatments that are compared to the baseline in the $i$th study and $\mathcal{T}^{-b_i}_i[k]$ as the $k$th treatment belonging to $\mathcal{T}^{-b_i}_i$.  Then 
\begin{align*}
    \bm{\delta}_i = \left(\delta_{i, b_i\mathcal{T}^{-b_i}_i[1]}, \ldots, \delta_{i, b_i\mathcal{T}^{-b_i}_i[t_i]}\right)' & \sim \mbox{N}_{t_i}(\bm{d}_{1i}, \mathbf{S}),
\end{align*}
where 
\begin{align*}
\bm{d}_{1i} & = \left(d_{1\mathcal{T}^{-b_i}_i[1]} - d_{1b_i}, \ldots, d_{1\mathcal{T}^{-b_i}_i[t_i]} - d_{1b_i}\right)';\\ 
\mathbf{S} & = \gamma\tau^2(\mathbf{J}_{t_i} - \mathbf{I}_{t_i}) + \tau^2\mathbf{I}_{t_i}.
\end{align*}
Here, $\mathbf{J}_{t_i}$ is a $t_i \times t_i$ matrix of ones and $\mathbf{I}_{t_i}$ a $t_i$-dimensional identity matrix.  

Sampling from the full conditionals of $\mu_{i, b_i}$, $\delta_{i, b_ik}$, and $\tau$ is done using a random walk Metropolis algorithm and in all cases we use a Gaussian candidate density. Then, we update each of the parameters by iterating on an individual basis through the following unnormalized full conditionals.
\begin{itemize}
  \item[-] Update $\mu_{i, b_i}$ based on 
    \begin{align*}
       [\mu_{i, b_i}|-] \propto  \prod_{k \in \mathcal{T}_i} \mbox{Binomial}(y_{ik}; n_{ik}, p_{ik})\mbox{N}(\mu_{i, b_i}; m_b, s_b),
    \end{align*}
    where ${\rm logit}(p_{ik})  = \mu_{i,b_i} + \delta_{i, b_ik}\mathbb{I}(b_i \neq k)$.
    
  \item[-] Update $\delta_{i, b_ik}$ based on
    \begin{align*}
       [\delta_{i, b_ik}|-] \propto  \prod_{k \in \mathcal{T}_i} \mbox{Binomial}(y_{ik}; n_{ik}, p_{ik}) \mbox{N}_{t_i}(\bm{\delta}_i; \bm{d}_{1i}, \mathbf{S}),
    \end{align*}
    where ${\rm logit}(p_{ik})  = \mu_{i,b_i} + \delta_{i, b_ik}\mathbb{I}(b_i \neq k)$.
    
  \item[-] Update $\tau$ based on
    \begin{align*}
       [\tau^2|-] \propto  \prod_{i=1}^N\mbox{N}_{t_i}(\bm{\delta}_i; \bm{d}_{1i}, \mathbf{S})\mbox{LogN}(\tau^2; m_{\ell}, s_{\ell}).
    \end{align*}
    
\end{itemize}
Updating $d_{1k}$ ($k = 2, \ldots, K$) and the other model parameters is slightly different for the three models.  We provide details for each model separately.
\begin{itemize}
  \item[] {\bf Gaussian effects Model.} Even though the full conditional $[d_{1k} | - ]$ is available in closed form, it is somewhat complex. Thus, we employ a Metropolis step to update $d_{1k}$.
    \begin{itemize}
        \item Update $d_{1k}$ using a Metropolis step and the following un-normalized full conditional
     \begin{align*}
        [d_{1k} | -] \propto \prod_{\substack{i: k \in \mathcal{T}_i, \\ c_k = h}} \mbox{N}_{t_i}(\bm{\delta}_i, \bm{d}_{1i}, \mathbf{S}) \mbox{N}(d_{1k}; m_d, s_d).
    \end{align*}
    \end{itemize}

  \item[] {\bf DP Gaussian effects Model.} Based on the component label representation of this model (see Section~4.1), an MCMC algorithm cycles through the following.
    \begin{itemize}
      \item[-]  Update $d^{\star}_h$  using a Metropolis step and the following un-normalized full conditional of $d^{\star}_h$
        \begin{align*}
          [d^{\star}_h | -] \propto \prod_{\substack{i: k \in \mathcal{T}_i, \\ c_k = h}} \mbox{N}_{t_i}(\bm{\delta}_i, \bm{d}^{\star}_{i}, \mathbf{S}) \mbox{N}(d^{\star}_h; m_d, s_d),
        \end{align*}
        where  $\bm{d}_{1i}$ and $\bm{d}^{\star}_i$ are connected by way of
       \begin{align*}
         \bm{d}^{\star}_i & = \left(d^{\star}_{c_{\mathcal{T}^{-b_i}_i(1)}} - d^{\star}_{c_{b_i}}, \ldots, d^{\star}_{c_{\mathcal{T}^{-b_i}_i(t_i)}} - d^{\star}_{c_{b_i}} \right)'  \\
          & = \left(d_{1\mathcal{T}^{-b_i}_i(1)} - d_{1b_i}, \ldots, d_{1\mathcal{T}^{-b_i}_i(t_i)} - d_{1b_i}\right)' \\
          & = \bm{d}_{1i}.
       \end{align*}

    \item[-] Update $c_k$ using 
      \begin{align*}
        \Pr(c_k = h | - ) \propto \pi_h\prod_{\substack{i: b_i = k, \\ c_{b_i} = h}} \mbox{N}_{t_i}(\bm{\delta}_i, \bm{d}^{\star}_{i}, \mathbf{S}) \prod_{\substack{i: k \in \mathcal{T}_i^{-b_i}, \\ c_{\mathcal{T}^{-b_i}_i[k]} = h}} \mbox{N}_{t_i}(\bm{\delta}_i, \bm{d}^{\star}_{i}, \mathbf{S}).
      \end{align*}

    \item[-] Update $V_h$ by way of the following full conditional
      \begin{align*}
        [V_h |-] \sim \mbox{Beta}(a^*, b^*),
      \end{align*}
      where $a^* = 1 + \sum_{k=2}^KI[c_k = h]$ and $b^* = \alpha + \sum_{k=2}^KI[c_k > h]$    
    \end{itemize}

  \item[] {\bf DP Spike-Slab Model.} Based on the component label representation of this model (see Section~4.2), an MCMC algorithm cycles through the following. 
    \begin{itemize}
      \item[-]  Update $d^{\star}_h$ using the same step as above. Namely, using a Metropolis step, sample from the following un-normalized full conditional of $d^{\star}_h$
        \begin{align*}
          [d^{\star}_h | -] \propto \prod_{\substack{i: k \in \mathcal{T}_i, \\ c_k = h}} \mbox{N}_{t_i}(\bm{\delta}_i, \bm{d}^{\star}_{i}, \mathbf{S}) \mbox{N}(d^{\star}_h; 0, v_0/3)^{s_h}~\mbox{NLP}(d^{\star}_h; p, 1, 1)^{1-s_h}.
         \end{align*}
      \item[-]  Update $c_k$ using the same arguments as above. Namely,
        \begin{align*}
          \Pr(c_k = h | - ) \propto \pi_h\prod_{\substack{i: b_i = k, \\ c_{b_i} = h}} \mbox{N}_{t_i}(\bm{\delta}_i, \bm{d}^{\star}_{i}, \mathbf{S}) \prod_{\substack{i: k \in \mathcal{T}_i^{-b_i}, \\ c_{\mathcal{T}^{-b_i}_i[k]} = h}} \mbox{N}_{t_i}(\bm{\delta}_i, \bm{d}^{\star}_{i}, \mathbf{S}).
        \end{align*}
        
      \item[-] Update $V_h$ using the same arguments as above. Namely,
        \begin{align*}
          [V_h |-] \sim \mbox{Beta}(a^*, b^*),
        \end{align*}
        where $a^* = 1 + \sum_{k=2}^KI[c_k = h]$ and $b^* = \alpha + \sum_{k=2}^KI[c_k > h]$    
        
      \item[-] Update $s_h$ using the following full conditional,
        \begin{align*}
        \Pr(s_h = 0 | - ) & \propto (1-\omega_0) \mbox{NLP}(d^{\star}_h ; p, 1, 1); \\ 
        \Pr(s_h = 1 | - ) & \propto (\omega_0) \mbox{N}(d^{\star}_h ; 0, v_0/3).
       \end{align*}
        
      \item[-] Update $\omega_0$ using the following full conditional,
       \begin{align*}
        [\omega_0 | - ] \sim \mbox{Beta}(a_{\omega} + \sum\mathbb{I}[s_h = 1],  b_{\omega} + \sum\mathbb{I}[s_h = 0]).
       \end{align*}
    \end{itemize}

\end{itemize}

\section{Additional Simulation Study Details}
\begin{figure}[hbt]
  \begin{center}
   \includegraphics[width=.7\textwidth]{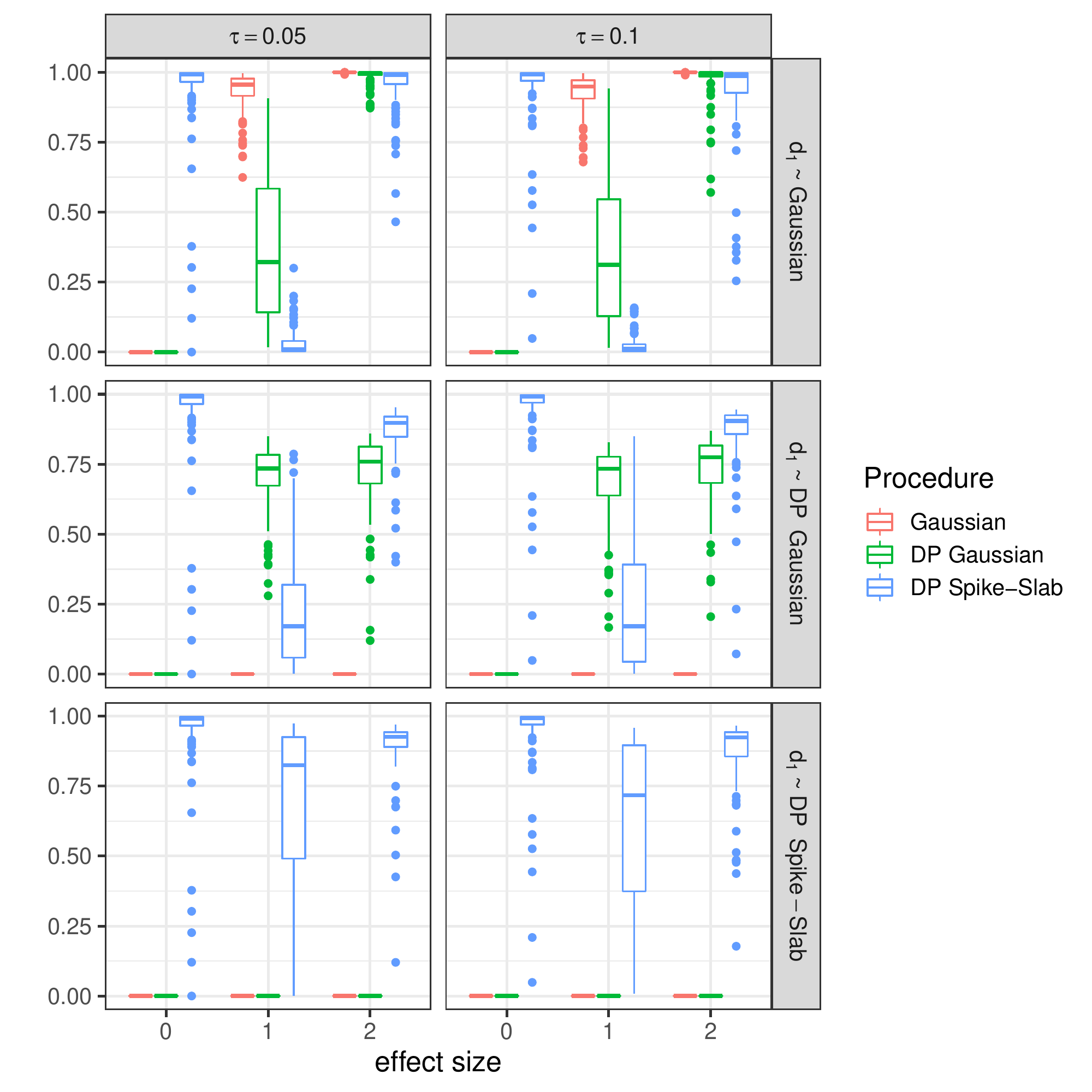}
 \end{center}
\caption{Posterior probability associated with the entire pairwise comparison graph that was used to generate data. This graph is comprised of 15 pairwise comparisons; as a result, the probabilities displayed are joint probabilities in the sense that they are associated with the probability of 15 pairwise comparisons simultaneously occurring.}
 \label{fig:graph_prob}
\end{figure}

%\bibliographystyle{agsm}

%\bibliography{../reference}